\documentclass[useAMS,usenatbib]{mn2e} 
\usepackage{setspace}
\usepackage{amsmath,amssymb,amsfonts}
\usepackage{graphicx}
\usepackage{epsfig}
\usepackage{epstopdf}
\usepackage{deluxetable}
\usepackage{pdflscape}
\usepackage{natbib}
\usepackage[T1]{fontenc}
\usepackage[caption=false]{subfig}
\usepackage{float}
\usepackage{multirow}

\usepackage{color}
\title[GC systems of the Leo II group] {The SLUGGS survey\footnotemark[0]\thanks{http://sluggs.swin.edu.au/}: Exploring the globular cluster systems of the Leo II group and their global relationships }
\author[S. S. Kartha et al.]{Sreeja S. Kartha$^{1}$ \thanks{E-mail: skartha@astro.swin.edu.au}, Duncan A. Forbes$^{1}$ \thanks{dforbes@astro.swin.edu.au}, Adebusola B. Alabi$^{1}$, Jean P. Brodie$^{2}$, 
 \newauthor Aaron J. Romanowsky$^{2,3}$, Jay Strader$^{4}$, Lee R. Spitler$^{5,6}$, Zachary G. Jennings$^{2}$,  \newauthor Joel C. Roediger$^{7}$\\
 $^{1}$ Centre for Astrophysics \& Supercomputing, Swinburne University, Hawthorn VIC 3122, Australia\\
 $^{2}$ University of California Observatories, 1156 High St., Santa Cruz, CA 95064, USA\\
 $^{3}$ Department of Physics and Astronomy, San Jos\'{e} State University, One Washington Square, San Jose, CA 95192, USA\\
 $^{4}$ Department of Physics and Astronomy, Michigan State University, East Lansing, MI 48824, USA\\
 $^{5}$ Macquarie University, Macquarie Park, Sydney, NSW 2113, Australia \\
 $^{6}$ Australian Astronomical Observatory, PO Box 915, North Ryde, NSW 1670, Australia\\
 $^{7}$ NRC Herzberg Astronomy \& Astrophysics, Victoria, BC V9E 2E7, Canada}

\begin{document}
 
\date{Released 2016 February  04 }

\pagerange{\pageref{firstpage}--\pageref{lastpage}} \pubyear{2016}

\def\LaTeX{L\kern-.36em\raise.3ex\hbox{a}\kern-.15em
    \kern-.1667em\lower.7ex\hbox{E}\kern-.125emX}

\newtheorem{theorem}{Theorem}[section]

\maketitle

\label{firstpage}

\begin{abstract}

We present an investigation of the globular cluster (GC) systems of NGC 3607 and NGC 3608 as part of the ongoing SLUGGS survey. We use wide-field imaging data from the Subaru telescope in the {\it g, r,} and {\it i}  filters to analyse the radial density, colour and azimuthal distributions of both GC systems. With the complementary kinematic data obtained from the Keck II telescope, we measure the radial velocities of a total of 81 GCs.

Our results show that the GC systems of NGC 3607 and NGC 3608 have a detectable spatial extent of $\sim$ 15, and 13 galaxy effective radii, respectively. Both GC systems show a clear bimodal colour distribution. We detect a significant radial colour gradient for the GC subpopulations in both galaxies. NGC 3607 exhibits an overabundance of red GCs on the galaxy minor axis and NGC 3608 shows a misalignment in the GC subpopulation position angles with respect to the galaxy stellar component.

With the aid of literature data, we discuss several relationships between the properties of GC systems and their host galaxies. A one-to-one relation between the ellipticities of red GCs and the galaxy stellar light emphasises the evolutionary similarities between them. In our sample of four slowly rotating galaxies with kinematically decoupled cores, we observe a higher ellipticity for the blue GC subpopulation than their red counterparts. Also, we notice the flattening of negative colour gradients for the blue GC subpopulations with increasing galaxy stellar mass. Finally, we discuss the formation scenarios associated with the blue GC subpopulation.

\end{abstract}


\section{Introduction}
To better understand the formation of early-type galaxies (ETGs), it is useful to study their oldest stellar components, such as globular clusters (GCs). GCs are mostly old ($\sim$10 Gyr, \citealt{Strader2005}), luminous ($\ge$ 10$^5$ L$_\odot$, \citealt{Brodie2011}) and compact (R$_e$ $\sim$ 3pc) star clusters. Establishing connections between the properties of GC systems and their host galaxies can help in understanding the formation of GC systems and hence, their parent galaxies. 

In almost all massive galaxies, GC systems are found to be bimodal in colour \citep{Larsen2001,Peng2006,Kim2013,Hargis2014}. Transforming colours to metallicity connects this bimodality with two stages of GC formation. The colour/metallicity distribution peaks are represented by blue/metal-poor and red/metal-rich GC subpopulations \citep{Brodie2012}. In a recent work, \citet{Forbes2015} determined the mean ages of blue and red GC subpopulations as 12.2 $-$ 12.8 and 11.5 Gyr respectively, suggesting that both subpopulations are very old. However, the two peaks differ in colour (e.g. 0.8 and 1.1 in ({\it g$-$i})) and metallicity (i.e. ([{\it Z/H}] peaks at $-$1.5 and $-$0.4 dex) values. Other properties of the two subpopulations differ such as azimuthal distribution, spatial distribution, radial colour distribution \citep{Strader2011,Park2013} and also kinematics \citep{Pota2013}.  Three scenarios have been suggested to explain the formation of these two distinct GC subpopulations: major-merger \citep{Ashman1992}, multi-phase collapse \citep{Forbes1997} and accretion \citep{Cote1998,Cote2000}. See \citet{Brodie2006} and \citet{Harris2010} for reviews. Many cosmological simulations of hierarchical galaxy formation have been used to investigate the characteristics  of GC systems (physical, dynamical, chemical etc.) in ETGs \citep{Beasley2002,Bekki2005,Bekki2008,Muratov2010,Griffen2010,Tonini2013,Katz2013,Gnedin2014,Trenti2015}.  

In order to associate the formation of different GC subpopulations with galaxy formation events, we explore the radial and azimuthal distributions of both subpopulations and compare with galaxy stellar light. Generally, the radial distribution of blue GCs is found to be more extended than that of both the red GC subpopulation and the galaxy stellar light. The red GC subpopulation is similar in radial distribution to the galaxy light \citep{Bassino2006b,Strader2011}, whereas the profiles of  the blue GC subpopulation show similarities with the X-ray surface brightness profiles \citep{Forbes2012a}. The radially extended blue GC subpopulation residing in galaxy haloes suggests that they are very old stellar components formed early in an in-situ dissipative collapse galaxy formation scenario, or accreted later, into the galaxy outskirts, in galaxy build up by minor merger formation scenario.  The similarity of red GCs with the galaxy stellar light supports coeval formation, but their origin (from the enriched gas of a parent galaxy or from accreted gas) is not clear.

 \cite{Park2013} studied the azimuthal distribution of GC systems in 23 ETGs using the data from the Advanced Camera for Surveys (ACS) mounted on the Hubble Space Telescope ({\it HST}). They found that the ellipticities of the red GC subpopulation match the galaxy stellar light ellipticities with a one-to-one correspondence. They also found that blue GC subpopulations show a similar but less tight relation.  \citet{Wang2013} using the ETGs, from the ACS Virgo Cluster Survey (VCS), concluded that both red and blue GC subpopulations align in position angle with the galaxy stellar light, although in a weaker way for blue GC subpopulations. Several single galaxy studies concluded that the galaxy stellar light is mimicked by red GC subpopulation in position angle and ellipticity, but the blue GC subpopulation is differently distributed (e.g. NGC 720, NGC 1023: \citealt{Kartha2014}, NGC 5813: \citealt{Hargis2014}, NGC 4365: \citealt{Blom2012}).

The decreasing mean colour of GC subpopulations with increasing galactocentric radius is a key diagnostic observation \citep{Bassino2006a,Harris2009b,Arnold2011,Faifer2011,Forbes2011,Blom2012,Usher2013,Hargis2014}. More specifically, the degree of steepness allows us to distinguish between two different formation processes, dissipation and accretion/merger \citep{Tortora2010}. Recently, \citet{Forbes2011} studied the colour gradient for NGC 1407 GC subpopulations and found that both GC subpopulations have a steep negative gradient within $\sim$ 8.5 effective radii (R$_e$) and a constant colour to larger radii. They explained this colour trend as being compatible with two-phase galaxy formation \citep{Oser2010}. This implies that the inner GCs are formed during a dissipative collapse phase, whereas the outer GCs are acquired during late accretion/mergers. Thus, exploring the radial colour distribution can reveal clues about formation events that happened in the host galaxy's history.

The layout of this paper is as follows. A brief literature review of the target galaxies is presented in Section 1.1. Section 2 describes the observations, data reduction techniques and initial analysis of imaging and spectroscopic data. Section 3 and 4 present the GC selection techniques and methods used to select the GC systems of individual galaxies. A detailed analysis of various GC system distributions (radial density, colour and azimuthal) for the selected GC systems is presented in Section 5. In Section 6, we discuss connections between the characteristics of galaxy stellar light and GC systems followed, in Section 7, by the conclusion.

\subsection{NGC 3607 and NGC 3608 in the Leo II group}
Here, we focus on the GC systems of the Leo II group. NGC 3607 and NGC 3608 are the brightest ETGs in the Leo II group. NGC 3607 is a near face-on lenticular galaxy while NGC 3608 is an E1$-$2 elliptical galaxy. In the same system there is a third galaxy, NGC 3605, which is a low mass galaxy of E4$-$5 morphology.  Table \ref{data} presents the main characteristics of the three galaxies with NGC 3607 as the central galaxy in the group. NGC 3608 and NGC 3605 are situated at a distance of 6 arcmin north-east and 2 arcmin south-west from NGC 3607. \citet{Kundu2001a,Kundu2001} investigated the GC systems of 57 ETGs including NGC 3607 and NGC 3608 using {\it HST}/Wide-Field Planetary Camera 2 (WFPC2) data in  {\it V} and {\it I} filters. For these galaxies they detected 130 and 370 GCs, respectively, from single pointing imaging. They did not find a sign of a bimodal colour distribution in either galaxy. 

\begin{table*}
\caption{Basic data for the target galaxies: NGC 3605, NGC 3607 and NGC 3608. Right ascension and declination (J2000) are taken from NASA/IPAC Extragalactic Database (NED). The galaxy distance, effective radius, heliocentric velocity, position angle and ellipticity, for NGC 3607 and NGC 3608, are from \citet{Brodie2014}. For NGC 3605, the galaxy distance, effective radius and heliocentric velocity are taken from \citet{Cappellari2011} whereas the position angle and ellipticity are obtained from HyperLeda \citep{Paturel2003}. Total {\it V}-band magnitudes are obtained from \citet{de1991}. The extinction correction for the {\it V}-band is calculated from \citet{Schlegel1998}. The absolute magnitude is derived from the {\it V}-band magnitude, distance and the extinction  correction. }
\begin{tabular}{lcclcccccccr} 
\hline
Name & RA & Dec & Type & D &V$_T$&A$_v$& M$_v^{\sc T}$ &R$_e$ &PA & $\epsilon$& Vel  \\
     & (h:m:s) & ($^o$:$'$:$\arcsec$)& & (Mpc)  &(mag)& (mag) &(mag) &(arcsec)& ($^o$)&& (km/s)\\
\hline
NGC 3605  & 11:16:46.6 & +18:01:02  &E4$-$5  &20.1&12.15  & 0.07& $-$19.36& 17&19& 0.40&661\\ 
NGC 3607 & 11:16:54.6 &+18:03:06  & S0 & 22.2 & 9.89  &0.07 & $-$21.86 & 39&125&0.13 &942\\  
NGC 3608 & 11:16:58.9 &+18:08:55 & E1$-$2 & 22.3  &10.76  &0.07& $-$20.98  & 30&82 &0.20 &1226\\ 
\hline
\end{tabular} 
\label{data}
\end{table*}

With the same {\it HST}/WFPC2 data, \cite{Lauer2005} investigated the surface brightness profiles of NGC 3607 and NGC 3608. They mentioned that NGC 3607 contains a symmetric, old and tightly wrapped outer dusty disk to which a second disk is settling in a perpendicular direction. They explained this observation as an infall of gas directly to the centre of galaxy with no interaction with the outer disk. They also detected the remnants of a pre-existing dusty disk in NGC 3608. \citet{Terlevich2002} derived the ages of 150 galaxies using the spectral line indices and found 5.8, 3.6 and 10 Gyr ages for NGC 3605, NGC 3607 and NGC 3608 respectively. Later, \citet{Rickes2009} investigated the metallicity distribution, stellar population and ionised gas in NGC 3607 using long-slit spectroscopy. They found stellar components ranging in age from 1 to 13 Gyr between the centre and a 30 arcsec radius of the galaxy's centre. As part of the ATLAS$^{3D}$ survey, \citet{McDermid2015} determined the mass-weighted ages for NGC 3605, NGC 3607 and NGC 3608 as 8.1 $\pm$ 0.8, 13.5 $\pm$ 0.7 and 13.0 $\pm$ 0.7 Gyr respectively. Also, from the ATLAS$^{3D}$ survey \citet{Duc2015} studied these galaxies using the deep multi-band images from the Canada France Hawaii Telescope. They mentioned that NGC 3607 and NGC 3608 are interacting galaxies with the presence of weak dust lanes and ripples.

Based on the {\it ROSAT} data, two peaks were detected in diffuse hot X-rays on NGC 3607 and NGC 3608 \citep{Mulchaey2003}. They proposed that the two galaxies are undergoing a merger. Later, \citet{Forbes2006a} detected an extended diffuse X-ray emission around the Leo II group.

The GC systems of the ETGs at the centre of the Leo II group have not yet been studied using wide-field imaging data. As part of the SLUGGS (SAGES Legacy Unifying Globulars and GalaxieS) survey \citep{Brodie2014}, we obtained wide-field data in three optical filters covering the central region of the Leo II group using the Suprime-Cam instrument on the Subaru telescope. With the aid of the imaging and spectroscopic data, we aim to understand the properties of the GC systems associated with each galaxy. 

\begin{figure*}
\centering
 \includegraphics[scale=9]{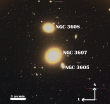} 
  \caption{A mosaic image showing the central 28 x 23 square arcmin area of the Leo II galaxy group. This Subaru/Suprime-Cam image is a combination of {\it g}, {\it r} and {\it i} filters. The target galaxies are labelled. The central galaxy, NGC 3607, is at a distance of 22.2 Mpc implying 1 arcsec = 0.107 kpc.  } 
\label{Galaxy}
\end{figure*}

\section{Data}
\subsection{Observations and reduction techniques}
Photometric data for the Leo II group was obtained using the Subaru Prime Focus Camera (Suprime-Cam; \citealt{Miyazaki2002}) mounted on the 8-meter Subaru telescope. The Suprime-Cam imager consists of ten CCDs with individual sizes of 2048 x 4096 pixels and a pixel scale of 0.202 arcsec, covering a maximum sky area of 34 x 27 square arcmin. Wide-field images were obtained during 2011 January 3 and 4. The sky conditions were good with an average seeing of  $\sim$ 0.81, 0.80 and 0.80 for the {\it g, r, i} filters, respectively. Multiple exposures in a dithered pattern were taken to fill the gaps between individual CCDs. In {\it g}, {\it r} and {\it i} filters the total exposure times were 3743, 1560 and 1200 sec, respectively. 

The individual exposures were reduced and combined using the Suprime-Cam Deep Field Reduction package 2 (SDFRED2; \citealt{Ouchi2004}) for each of the three filters. The pre-processing of images included flat fielding, distortion corrections and atmospheric dispersion corrections. The pipeline also features custom-made modifications to improve the sky subtraction and alignment between multiple exposures. We employed SExtractor \citep{Bertin1996}  \footnote{http://www.astromatic.net/software/} and Montage \footnote{http://montage.ipac.caltech.edu/index.html} for the alignment process.  All point sources three sigma above the background level are identified on each CCD image using SExtractor. The positions of these point sources are matched with a reference catalogue (here we use the Sloan Digital Sky Survey) to create an astrometric solution. The astrometric solution is used by the Montage program to align and combine the individual images, generating mosaic images in the three filters. A combination of {\it g}, {\it r} and {\it i} filter mosaic images is shown in Figure \ref{Galaxy}.

We also obtained a single pointing covering the central region of NGC 3607 from the Hubble Legacy Archive (HLA). This was taken in the F814W ({\it I}) filter using the ACS instrument. The Wide-Field Channel  on the ACS consists of two 2048 x 4096 CCDs with a 0.049 arcsec pixel scale, and 3.37 x 3.37 square arcmin field of view. A custom-made pipeline (for detailed explanation see \citealt{Spitler2006}) is employed to reduce the ACS data. The pipeline provides source positions and half light radii for all the detected sources, which are utilised for a preliminary selection of GCs in the Subaru/Suprime-Cam imaging (see Section 3). 

Complementary spectroscopic data were obtained using the DEep Imaging Multi-Object Spectrograph (DEIMOS, \citealt{Faber2003}) on the Keck II telescope. The field of NGC 3607 was targeted on five nights during 2013 January 10 -- 12 and 2014 January 26 and 27 as part of the SLUGGS survey.  We used five slit-masks for good azimuthal coverage and the seeing per night was 0.87 $\le$ FWHM $\le$ 1.15 arcsec with a total exposure time of $\sim$10 hours.  DEIMOS was used with 1200 l/mm grating centered on 7800 \AA, with slit widths of 1 arcsec. In this way, we have a wavelength coverage from 6500 -- 8700 \AA~ and spectral resolution of $\sim1.5$ \AA. We reduced the raw spectra using the \texttt{IDL SPEC2D} reduction pipeline together with dome flats and arc lamp spectra. The pipeline produces sky-subtracted GC spectra that covers the CaT absorption lines in the near-infrared (8498, 8542, 8662 \AA) and H$\alpha$ line at 6563 \AA (where possible). 

We obtain the radial velocities from our science spectra using the \texttt{FXCOR} task in \texttt{IRAF} by measuring the doppler shift of the CaT lines, cross-correlating each Fourier transformed science spectrum with the Fourier transformed spectra of 13 Galactic template stars. In practice, we require that the strongest CaT lines (8542, 8662 \AA) be present and where possible the H$\alpha$ line as well. Where the lines are not properly defined, but the velocity is consistent with either galaxy, the GC is classified as marginal. Objects with velocities less than 350 km/s are classified as Galactic stars and those with velocities greater than 1800 km/s as background galaxies. Our final catalogue has 75 GCs and 7 ambiguous objects (see Table \ref{vel_table} in Appendix \ref{appB}). Here, 'ambiguous' denotes that either the velocity or position has a mismatch with the target galaxies, but it has confirmed characteristics of a GC.

\subsection{Photometry}
Before carrying out any photometric analysis, the galaxy light was subtracted in each of the three mosaic images. The two large galaxies are individually modelled using {\tt IRAF} task {\tt ELLIPSE} with the center of the galaxy, the major axis position angle (PA) and the ellipticity ($\epsilon$) as free fitting parameters. During the fitting process the bright stars were masked before modelling the galaxy light. The best fit galaxy model produces radial profiles of surface brightness, position angle and ellipticity measurements for both the galaxies. We made use of galaxy light subtracted images to improve the source detection in the central regions of target galaxies. 

We utilised SExtractor for source identification and photometry.  We instructed SExtractor to identify a probable source only if it has a minimum of 5 adjacent pixels with  a flux higher than three sigma above the local background. SExtractor estimates the total instrumental magnitudes for the detected sources using Kron radii \citep{Kron1980} in the automatic aperture magnitude mode.  It provides an output list of point sources with position and magnitude. As standard stars were not observed for zeropoint calibration, we exploited the bright stars ({\it i} $<$ 22) present in the galaxy field. A match between these bright stars and the Sloan Digital Sky Survey catalogue (data release 7 version) was used for the flux calibration in all three mosaic images. Photometric zeropoint magnitudes in three filters are derived from the best-fit linear relationship between the instrumental magnitudes of bright stars and calibrated magnitudes from the SDSS catalogue. Estimated zeropoints in {\it g, r, i} bands are 28.68 $\pm$ 0.08,  28.92 $\pm$ 0.12, 28.78 $\pm$ 0.15 magnitudes, respectively. All magnitudes have had the zeropoint correction applied. The galaxy photometry is corrected for Galactic extinction using the dust extinction maps from \citet{Schlegel1998}. Hereafter, all the magnitudes and colours cited are extinction corrected.

\section{Globular cluster selection}
\label{GCSelect}
The large  galaxies, NGC 3607 and NGC 3608, are at an assumed distance of 22.2 $\pm$ 0.1 Mpc \citep{Brodie2014} and NGC 3605 taken to be 20.1 Mpc \citep{Cappellari2011}. For GC identification, a match of object positions between the three bands is carried out at first, in order to eliminate the false detections. Afterwards a separation between extended objects (galaxies) and point source objects (both GCs and stars) is incorporated. This separation is based on the surplus light detected beyond the extraction aperture. Objects showing large difference between the extraction aperture and an outer aperture are considered as extended sources and are removed (see \citealt{Kartha2014} for details). 

We employ a colour-colour selection as the next step to identify the GC candidates. To aid this selection, we used the position and half light radius of the sources from the {\it HST}/ACS data.  An upper limit of $\sim$ 9 pc at the distance of NGC 3607, for GC candidature is applied, and the selected objects are visually verified. A positional match between the Subaru objects and the GCs selected on the {\it HST}/ACS image is carried out and then the half light radius is attached to the Subaru list for the common objects. Hence we create a list of probable GCs with their positions, three magnitudes from the Subaru/Suprime-Cam data, and half light radii from the {\it HST}/ACS data. From earlier studies, e.g., figure 6 in \citet{Blom2012} and figure 3 in \citet{Pota2013}, it is evident that the GCs populate a particular region in the colour-colour diagram. With the above list we identify the locus of GCs in ({\it {r$-$i}}) verses ({\it g$-$i}) colour space, implementing similar procedures as explained in \citet{Spitler2008} and \citet{Blom2012}. The GC candidates, along with neighbouring objects showing a 2$\sigma$ deviation from the selected region, are chosen as final GC candidates. The selected GCs range over 0.6 $<$ ({\it g$-$i}) $<$ 1.4, which corresponds to a metallicity range of $-$1.94  $<$ [{\it Z/H}] $<$ 0.86 using the empirical relation given in \citet{Usher2012}. The upper and lower cut off in {\it i} band magnitude are 20.4 and 24.4 magnitudes, respectively. At the distance of NGC 3607 objects brighter than 20.4 magnitude include ultra compact dwarfs \citep{Brodie2011} while the lower limit is one magnitude fainter than the turnover magnitude for the GC system. This final list of GC candidates include $\sim$ 1000 objects from NGC 3605, NGC 3607 and NGC 3608.

\section{Defining the GC systems of each galaxy}
We derive the stellar mass of NGC 3605 as log(M$_\star$) = 10.76 M$_{\odot}$ from the galaxy {\it V}-band magnitude (see Table \ref{data}) and the mass to light ratio from \citet{Zepf1993}. The extent of the GC system of NGC 3605 is calculated from the stellar mass in the empirical relation between GC system extent and the galaxy stellar mass (equation 7 in \citealt{Kartha2014}).  A GC system extent of $\sim$ 40 arcsec is derived from the calculation and we assume a maximum of 1 arcmin extent for NGC 3605. We detect 10 objects in the 1 arcmin region around NGC 3605 and eliminate them from the following calculations. The surface density distribution of GCs around NGC 3605 has been investigated and we find a constant GC density, implying no contamination from NGC 3605 to the NGC  3607 or NGC 3608 GC systems.
  
The remaining GC candidates are a combination of objects from NGC 3607 and NGC 3608. In order to classify their individual GC systems, we invoke two methods, based on surface brightness and position angle of the host galaxies.

\subsection{Surface brightness method}
\label{SB}

\begin{figure}
\centering
\includegraphics[scale=0.5]{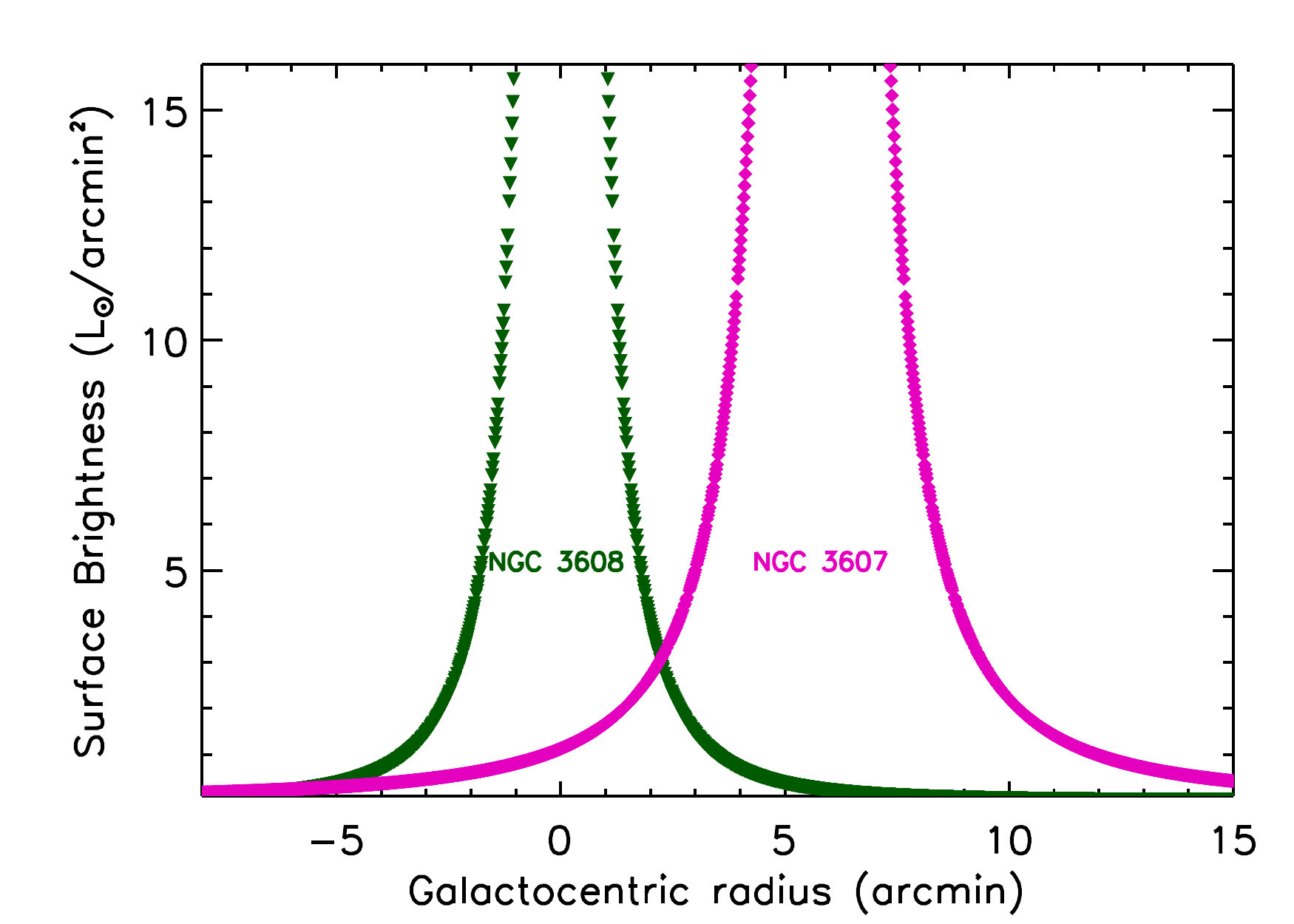}
 \caption{ Surface brightness profiles for individual galaxies. The {\it i}- band profiles have been extracted using \texttt{IRAF ELLIPSE} task and extrapolated towards larger radii from the centres of NGC 3607 and NGC 3608. The negative to positive radius represents the declination axis centered on NGC 3608.} 
\label{SB}
\end{figure}

\begin{figure}
\centering
\includegraphics[scale=0.45]{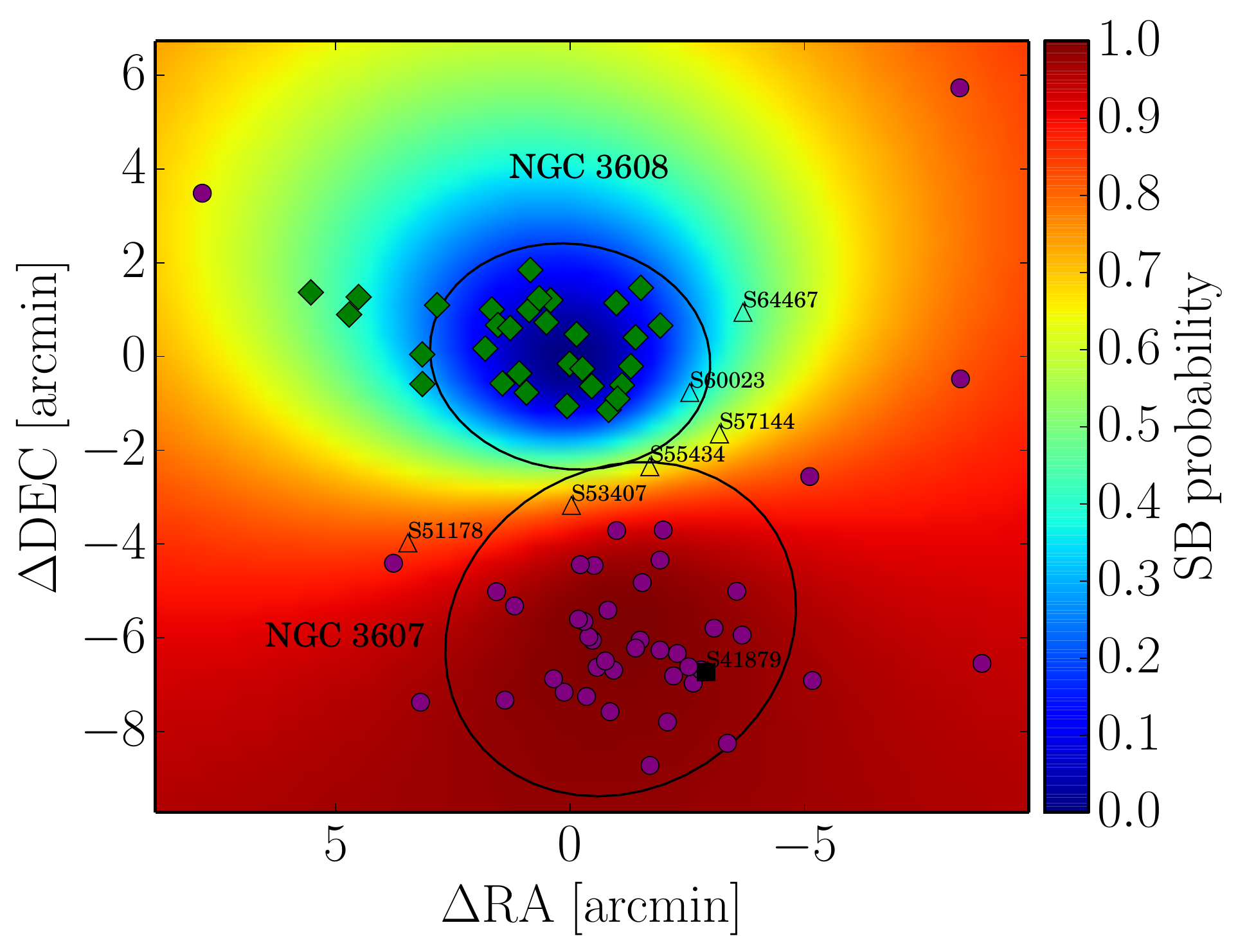}
 \caption{Spectroscopically confirmed GCs of NGC 3607 and NGC 3608. The galaxy centres for NGC 3608 and NGC 3607 are, respectively, at co-ordinates (0,0) and ($-$1.1,$-$5.8). The magenta circles and green diamonds represent the GC members of NGC 3607 and NGC 3608, while open triangles and black square represent ambiguous GCs (with IDs denoted) and one extreme object (ID: S41879). The colour map in the background represents  the membership probability from the surface brightness and the colour coding is shown to the right. The black ellipses represent six effective radii for the two galaxies with their respective ellipticity and position angle of the galaxy stellar light (refer to Table \ref{data}). }
 \label{vel_map}
\end{figure}

The galaxy light for both galaxies is modelled and extracted using the \texttt{IRAF} task \texttt{ELLIPSE}. The individual surface brightness profiles are fit with S\'{e}rsic profiles \citep{Graham2005}. We extrapolate these profiles to larger galactocentric radius ($\sim$ 15 arcmin) and use these extrapolated profiles to represent the stellar light profiles of individual galaxies to large radius.  Figure \ref{SB} shows the surface brightness profiles of NGC 3607 and NGC 3608. Based on the position of each GC, its membership probability is computed from the ratio of surface brightness of NGC 3607 to NGC 3608. Hereafter we refer to this as the surface brightness (SB) method. GCs with a probability greater than 55 percent are counted as members of NGC 3607, while less than 50 percent are classified as members of NGC 3608. The 6 R$_e$ ellipses overlap around 55 percent SB probability (see Figure \ref{vel_map}). We classify the GCs with probability between 55 and 50 percent as ambiguous objects.  

 \subsection{Major axis method}
\label{MA}
We employed a second method called the major axis (hereafter MA) method, to separate the GC systems of the two galaxies. In this method, we divided the GCs along the photometric major axis (125 and 82 degrees for NGC 3607 and NGC 3608, respectively) and selected the hemisphere pointing away from the other galaxy. Thus, the selection of GCs for NGC 3607 includes GCs in the position angles 125 to 305 degrees and for NGC 3608 GCs from 0 to 82 and 262 to 360 degrees. This method excludes the region of maximum tidal interaction between the two galaxies. \citet{Coccato2009}  adopted a similar method for disentangling the planetary nebulae (PNe) of NGC 3608. To eliminate the contaminants from NGC 3607, they excluded the PNe on the southern side of NGC 3608, which is equivalent to the major axis method used here.

\begin{figure}
\centering
\includegraphics[scale=0.5]{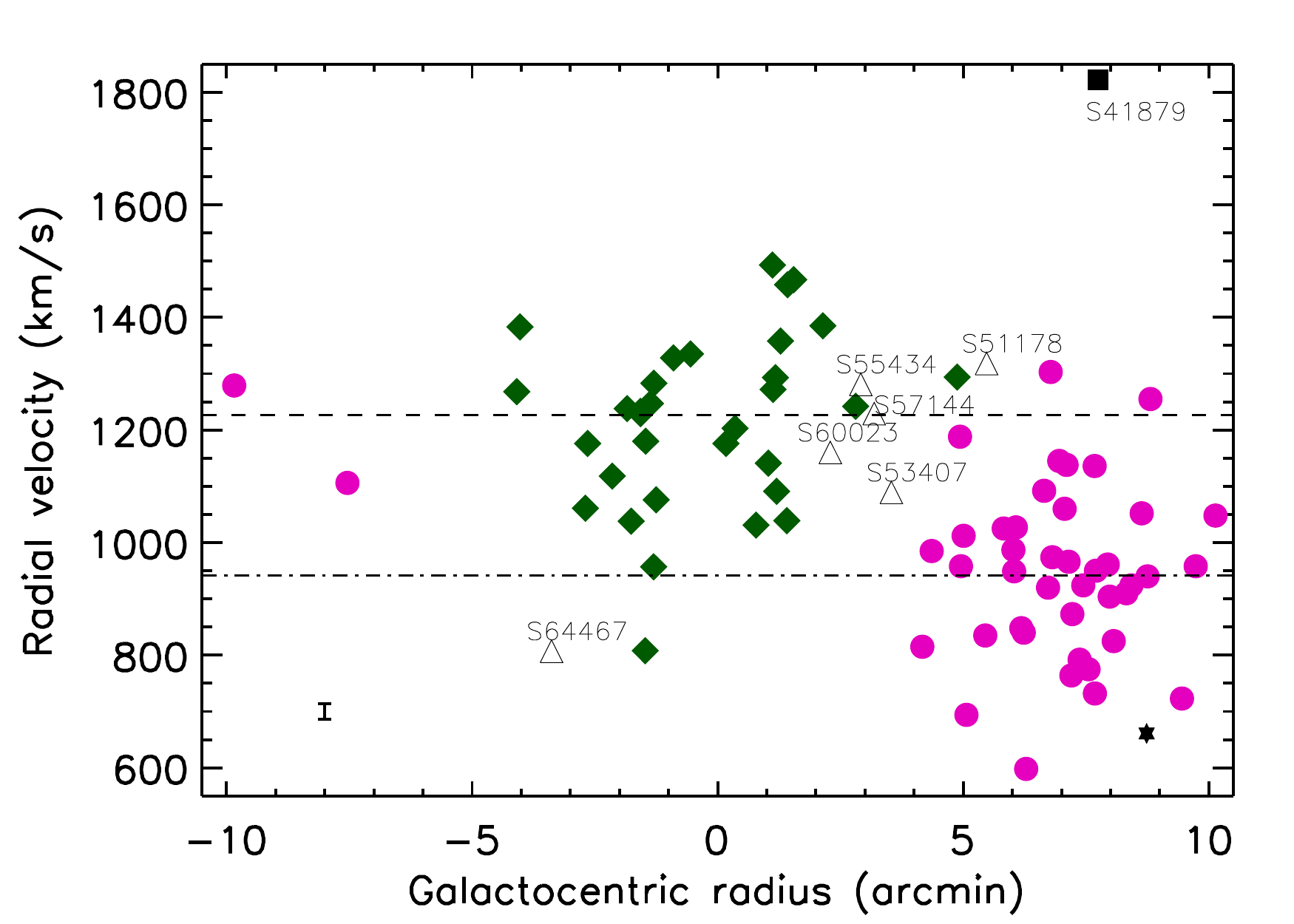}
 \caption{ Velocity distribution of spectroscopically confirmed GCs as a function of radius with respect to NGC 3608. The NGC 3607 and NGC 3608 members are represented with magenta circles and green diamonds, while  marginal GCs and one extreme object (ID: S41879) with open triangles and a filled square. The position of NGC 3605 is represented with a black star. The dot-dashed and the dashed horizontal lines represent the galaxy systemic velocities for NGC 3607 and NGC 3608, respectively. An average error of 14 km/s is shown at the lower left.}
 \label{vel_rad}
\end{figure}

\subsection{Analysis of kinematic data}
\label{kin_data}
We obtained the radial velocity measurements for 82 (confirmed plus marginal) GCs in the field of the Leo II group. The galaxy systemic velocities for NGC 3607 and NGC 3608 are 942 and 1226 km/s \citep{Brodie2014}, respectively. To assign the membership of GCs to individual galaxies, we performed a biweight estimator distribution (following \citealt{Walker2006}) based on the right ascension, declination and line of sight velocity of each GC. The GCs within 2$\sigma$ ($\sigma$ is the standard deviation calculated from the velocity distribution) from the central galaxy velocity are assigned membership to the corresponding galaxy, while keeping marginal members as velocities between 2$\sigma$ to 3$\sigma$.  Figure \ref{vel_map} displays positions of spectroscopically confirmed GCs on a SB probability map. The background map shows the SB probability used in the separation of GCs (see Section \ref{SB}).  The positions of individual galaxy GCs (as determined using velocities) fall on the same region derived from the SB method, confirming the robustness of the SB probability method for classifying the GCs.  The distribution gives  43 and 32 GCs, respectively, as NGC 3607 and NGC 3608 members. 

In addition, we classified the 7 ambiguous objects as 6 GCs and one extreme member. The extreme member S41879 has a velocity of 1822 $\pm$ 22 km/s, but positionally it is projected near the centre of NGC 3607 (see Figure \ref{vel_map}) in the 2D map.  Assuming it lies at the distance of NGC 3607 (D = 22.2 Mpc), then it has M$_ i$ = $-$9.97 mag. From the line of sight velocity and H$_0$ = 70 (km/s)/Mpc, we calculate the distance as 26  Mpc and hence the magnitude M$_i$ = $-$10.31 mag.  This suggests that it is a possible UCD (see \citealt{Brodie2011}). To confirm this, we checked the {\it HST} image for an estimation of its size.  Unfortunately, this object is placed in the central gap region of the {\it HST} pointing. We examined the Subaru image and found that the object is very circular in shape.  Another possibility is an intra-group GC, as it is blue ({\it g$-$i}) = 0.623, circular in shape and lies in the projected region between NGC 3607 and NGC 3605. With the above information, we suggest that this extreme object might be a background UCD or an intra-group GC.  Eliminating this extreme object, we have 81 spectroscopically confirmed GCs for NGC 3607 and NGC 3608.

\begin{figure}
\centering
\includegraphics[scale=0.37]{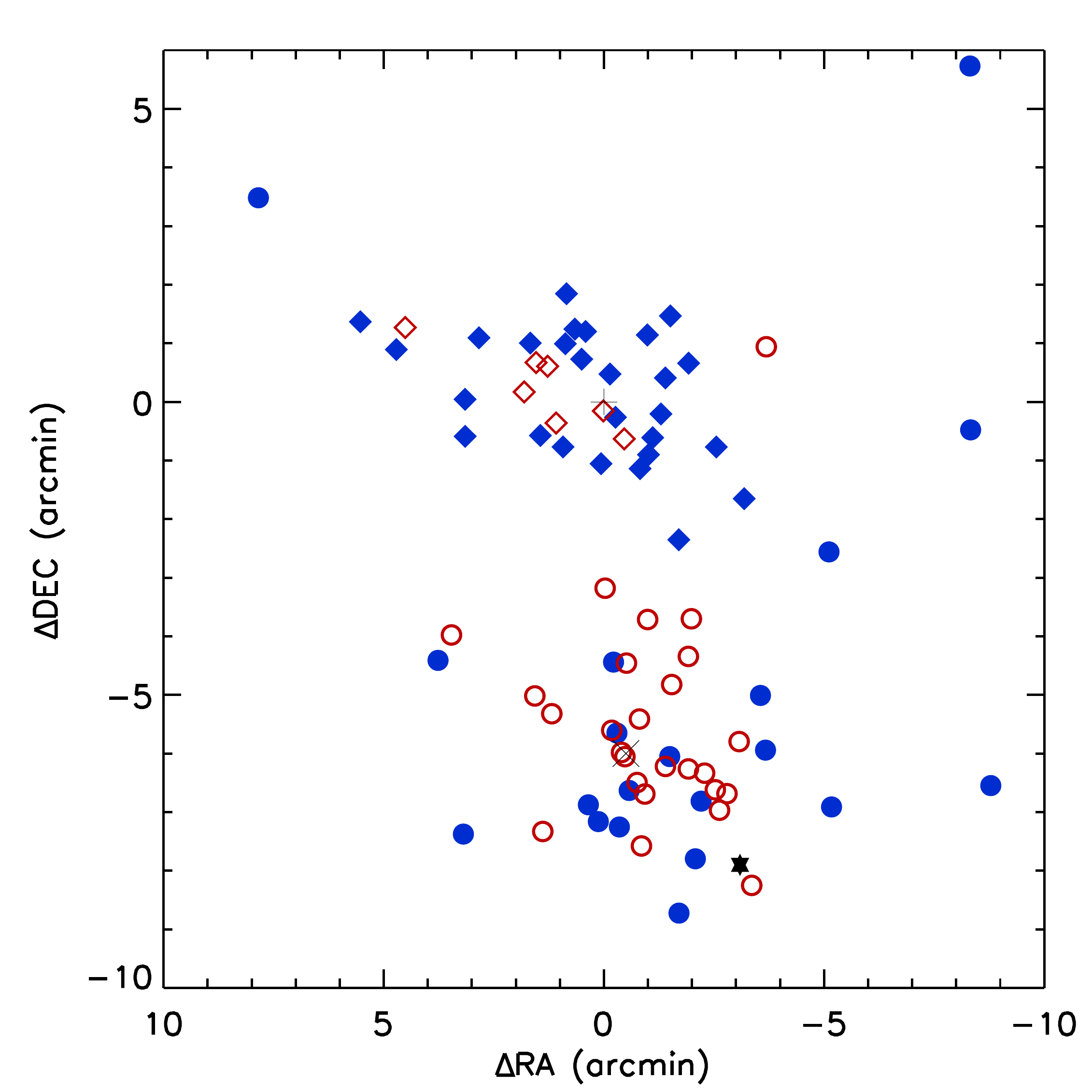}
 \caption{GC subpopulations for the spectroscopically confirmed GC systems of NGC 3607 and NGC 3608. The centre of NGC 3605, NGC 3607 and NGC 3608 are denoted with star ($-$3.1, $-$7.9), cross ($-$1.1,$-$5.8) and plus (0,0) symbols. The diamonds and circles (both open and filled) represent, respectively, the GCs of NGC 3608 and NGC 3607. The red and blue colours represent the blue and the red subpopulations for both galaxies. }
\label{col_vel_map}
\end{figure}

Figure \ref{vel_rad} shows the velocity distribution of GCs with galactocentric radius measured from the centre of NGC 3608. The six marginal GCs are labelled in Figures \ref{vel_map} and \ref{vel_rad}. Based on both these figures, we assign a membership for the marginal GCs. Note here that this manual membership assignment is unimportant for any broad conclusions of this study. S51178  is positionally close towards NGC  3607 with velocity > 1300 km/s. But according to the SB probability, this GC has > 80 percent probability to be associated with NGC 3607.  Hence, considering these facts we assign it to NGC 3607 as GC44 (name given in Table \ref{vel_table}). Based on the SB probability and velocity measurement, S53407 is assigned to NGC 3607 (GC45). The position of S64467 is close to NGC 3608 with 50 percent probability, but having a velocity of 807 km/s supports a membership with NGC 3607 (GC46). S60023 has a 70 percent probability with NGC 3608 and with a velocity of 1160 km/s. Hence, S60023 is a probable member of NGC 3608 (GC33). S55434 (GC34) and S57144 (GC35) are GCs with velocities 1281  and 1229  km/s, respectively. Both fall on the probability region of  $\sim$ 60 percent for NGC 3607. However, a membership to NGC 3608 is allocated for these GCs based on the positional closeness and velocities. Hence, S60023, S55434, S57144 are NGC 3608 members and S51178, S64467, S53407 are NGC 3607 members. Finally, NGC 3608 and NGC 3607  have 35 and 46 spectroscopically confirmed GCs, respectively. 

The mean velocities estimated from the GC systems of NGC 3607 and NGC 3608 are 963 and 1220 km/s,  respectively, in good agreement with galaxy central velocities. Estimates of the GC system velocity dispersions for NGC 3607 and NGC 3608  are 167 and 147 km/s, respectively. \citet{Cappellari2013} found central velocity dispersions of 206.5 $\pm$ 10 and 169.0 $\pm$ 9 km/s from the galaxy stars, respectively, for NGC 3607 and NGC 3608. 

\subsubsection{GC subpopulations}
Currently, we have 46 and 35 spectroscopically confirmed GCs, respectively, for NGC 3607 and NGC 3608. We have classified the GCs into blue and red subpopulations based on a constant colour division with galactocentric radius due to small number statistics. The GMM algorithm (explained in Section \ref{GCb}) gives a ({\it g$-$i}) dividing colour of 0.87 for NGC 3607 and 0.93 mag for NGC 3608 (from photometric measurements). We used these colours to separate the blue and the red subpopulations of the two galaxies as shown in Figure \ref{col_vel_map}. From the photometric analysis of the GC subpopulations, we obtained 62 and 38 percent blue and red subpopulations (see Section \ref{8subpop}), respectively. 

\section{Analysis of photometric data}
Below we describe the radial density, colour and azimuthal distributions of the NGC 3607 and NCG 3608 GC systems. Note here that the GC systems are selected from the colour-colour space discussed in Section \ref{GCSelect}. 

\begin{figure}
\centering
 \includegraphics[scale=0.5]{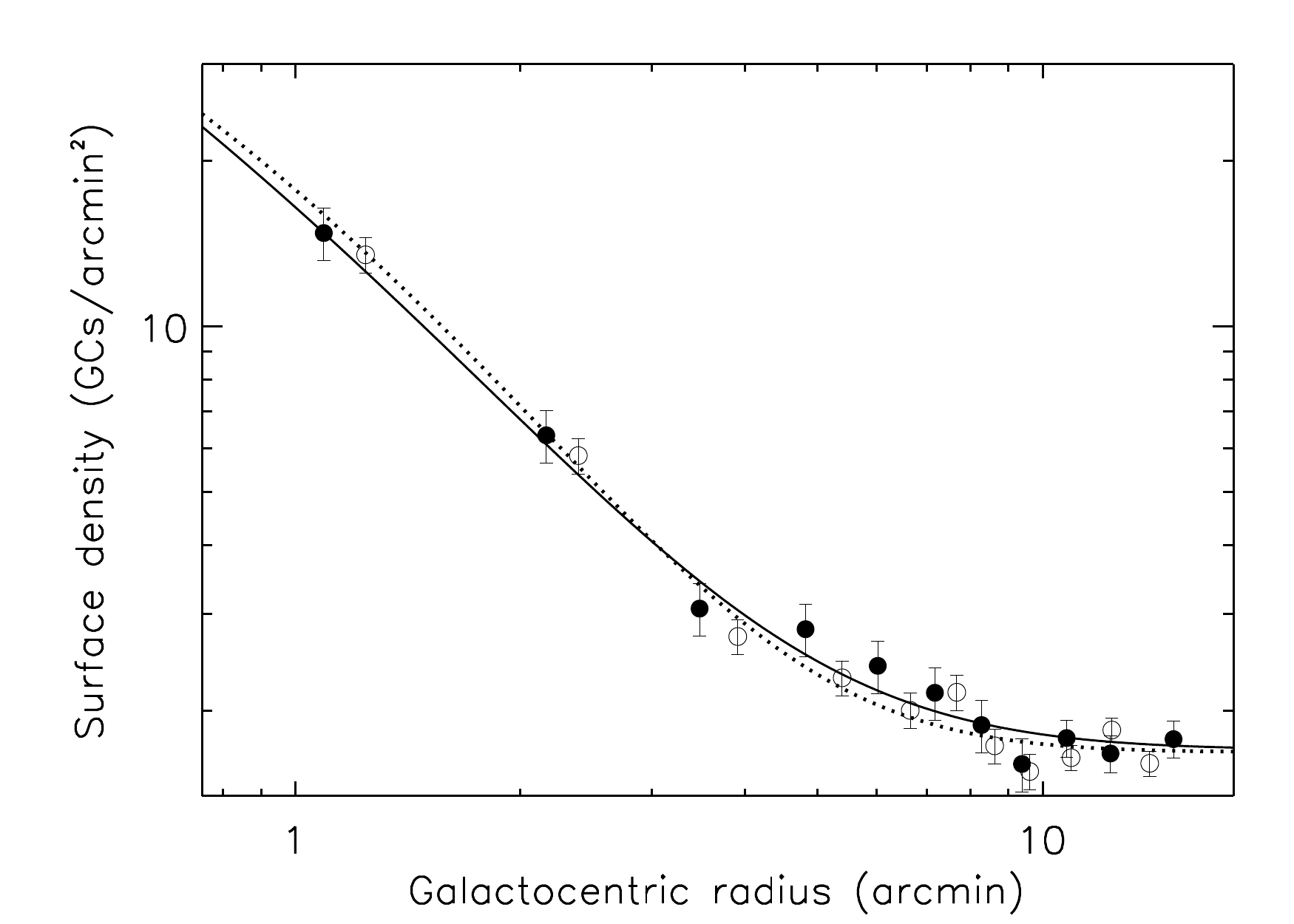} 
  \caption{ Surface density distribution for the GC system of NGC 3607. The GCs are selected  via the SB and  the MA methods shown by filled and open circles. The solid and the dotted lines represent the S\'{e}rsic fits for the GCs selected from each method. The GC system reaches the background around a galactocentric radius of 9.5 $\pm$ 0.6 arcmin, in agreement with the expected value using the galaxy stellar mass in the relation of \citet{Kartha2014}. } 
\label{SD_07}
\end{figure}

\subsection{GC system of NGC 3607}
\subsubsection{Radial density distribution}
\label{SD}
To derive the radial distribution of the GC system, we define radial bins up to a galactocentric radius of 16.9 arcmin. Then the effective area coverage is obtained for each radial annulus. The area is corrected for the presence of saturated stars and for any regions outside the detection area. The GC number in each annulus is then divided by the effective spatial area to determine the spatial density in that particular annulus.  The errors are calculated using Poisson statistics. 

 We obtained the GC system surface density using two methods. In the SB method (refer Section \ref{SB}), a correction is applied for the missing area due to NGC 3608 and NGC 3605. In the MA method (refer Section \ref{MA}), the number density is doubled in each radial bin. The radial density distribution is fitted with a combination of S\'{e}rsic profile plus a background parameter to estimate the effective radius and the background value for the GC system. The fitted surface density profile is:
\begin{equation}
N(R) = N_e ~exp \left[-b_n \left(\frac {R}{R_e}\right)^\frac{1}{n} - 1\right] + bg
\label{sersic}
\end{equation}
where {\it N$_e$} is the surface density of the GCs at the effective radius {\it R$_e$}, {\it n} is S\'{e}rsic index or the shape parameter for the profile, {\it b$_n$} is  given by the term 1.9992n $-$ 0.3271 and {\it bg} represents the background parameter. Note that the radius {\it R} is the centre of each radial bin.

Figure \ref{SD_07} shows the density profile of the GC system for NGC 3607 only. The GCs brighter than the turnover magnitude, {\it i} = 23.5, only are considered. The plot displays the density values derived from the two different methods, i.e. SB and MA methods. Both are fitted with the profile given in Equation \ref{sersic}. In the density distribution plot for NGC 3607, the SB and MA methods used 1170 and 907 objects, respectively. It is evident from the figure that both methods yield consistent results and the profile reaches the background at a galactocentric radius of 9.5 $\pm$ 0.6 arcmin (61 $\pm$ 3 kpc). 

\citet{Kartha2014} found an empirical relation between the galaxy stellar mass and the extent of its GC system. The relation is as follows:
\begin{equation}
\text{GCS~extent}~(\text{kpc}) = [(70.9~\pm~11.2) \times \text{log}(M/M_{\odot})]  - (762~\pm~127).
 \label{GCSextent}
\end{equation}
\begin{table}
\centering
\caption{Fitted parameters for the surface density profile of the NGC 3607 GC system. The first column represents the method used for deriving the surface density profile. The effective radius, the S\'{e}rsic index and the background estimation are given in the following three columns. The last column presents the extent of the GC system as measured. The error values given are 1-sigma uncertainties.}
\begin{tabular}{ccccc}
\hline
Method  & R$_e$ & n & bg  & GCS ext.\\ 
            & (arcmin) & & (arcmin$^{-2}$) & (arcmin) \\ 
\hline\hline
SB & 2.45 $\pm$ 0.54 & 2.74 $\pm$ 1.76  & 1.70 $\pm$ 0.15 &9.4 $\pm$ 0.6\\
MA & 1.99 $\pm$ 0.29 & 1.97 $\pm$ 1.19 &1.68 $\pm$ 0.08 & 9.6 $\pm$ 0.5\\ 
\hline
\end{tabular}
\label{surfden07}
\end{table}

NGC 3607, an S0 galaxy, with absolute {\it V}-band magnitude M$_V^T$ = $-$21.87 and assumed mass to light ratio of 7.6 (given in \citealt{Zepf1993})  has a host galaxy mass, log(M/M$_\odot$) = 11.56. The GC system extent for NGC 3607 determined using the above equation is 57 $\pm$ 3 kpc, in good agreement with the direct estimation using the wide-field Subaru/Suprime-Cam image (61 $\pm$ 3 kpc). 

\begin{figure}
\centering
\includegraphics[scale=0.33]{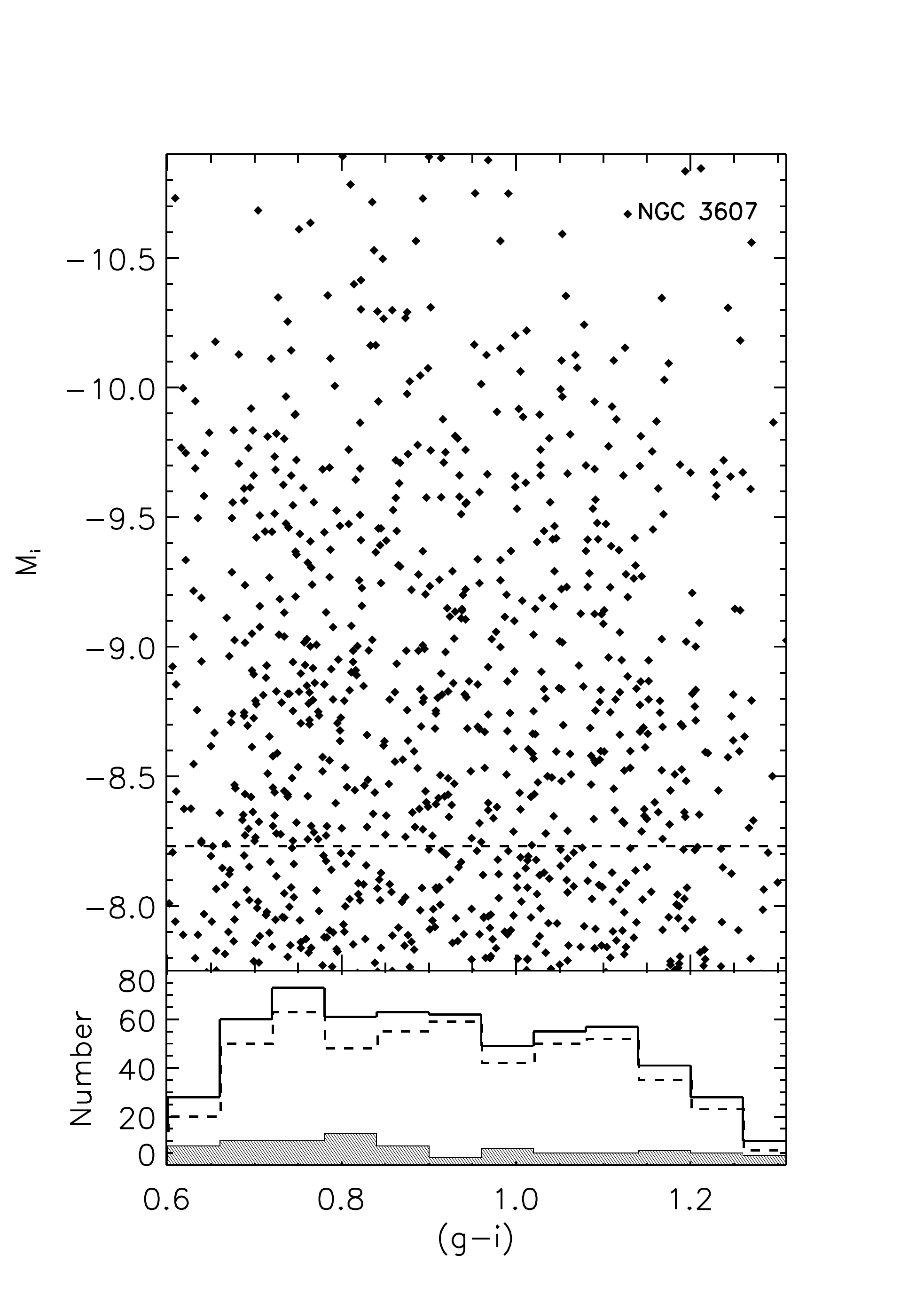}
\caption{ Colour magnitude diagram for NGC 3607. The top panel represents the GCs brighter than M$_i$ = $-$7.75 mag (0.5 fainter than the turnover magnitude) within the extent of NGC 3607 GC system. The dashed line represents the turnover magnitude in {\it i} filter, M$_i$ = $-$8.23 mag. The bottom panel represents the colour histogram of NGC 3607 GC system. The open, shaded and dashed histograms represent the GCs which are brighter than the turnover magnitude, the estimated background contamination and the background corrected colour histograms. }
\label{CMD_07}
\end{figure}

\subsubsection{GC bimodality}
\label{GCb}

 Figure \ref{CMD_07} shows the colour magnitude diagram of NGC 3607 GCs. The GCs, brighter than M$_i$ = $-$7.75 mag, within the GC system extent of NGC 3607 are shown in the diagram. The bottom panel contains the histogram of GCs which are brighter than the turnover magnitude (M$_i$ = $-$8.23 mag) along with the background contamination. To estimate the background contamination, we made use of the detected objects beyond the GC system extent of the galaxy. For NGC 3607, the objects beyond 11 arcmin (as GC system extent is 9.5 $\pm$ 0.6 arcmin) are considered as background contamination. We applied an areal correction, if needed. The background corrected colour histogram is also shown in Figure  \ref{CMD_07} and it includes 611 GCs.
 
 To quantify the colour distribution of the GC system, we used the gaussian mixture modeling (GMM, \citealt{Li2014, Muratov2010}) algorithm on the GC system ({\it g$-$i}) colour, after background correction.  The algorithm tests for a multimodal colour distribution over unimodal.  To be a significant multimodal GC system distribution, the following three statistics should be, 1. low values for the confidence level from the parametric bootstrap method, 2. the separation (D) between the means and the respective widths greater than 2, and 3. negative kurtosis for the input distribution. 
 
 \begin{figure}
\centering
 \includegraphics[scale=0.5]{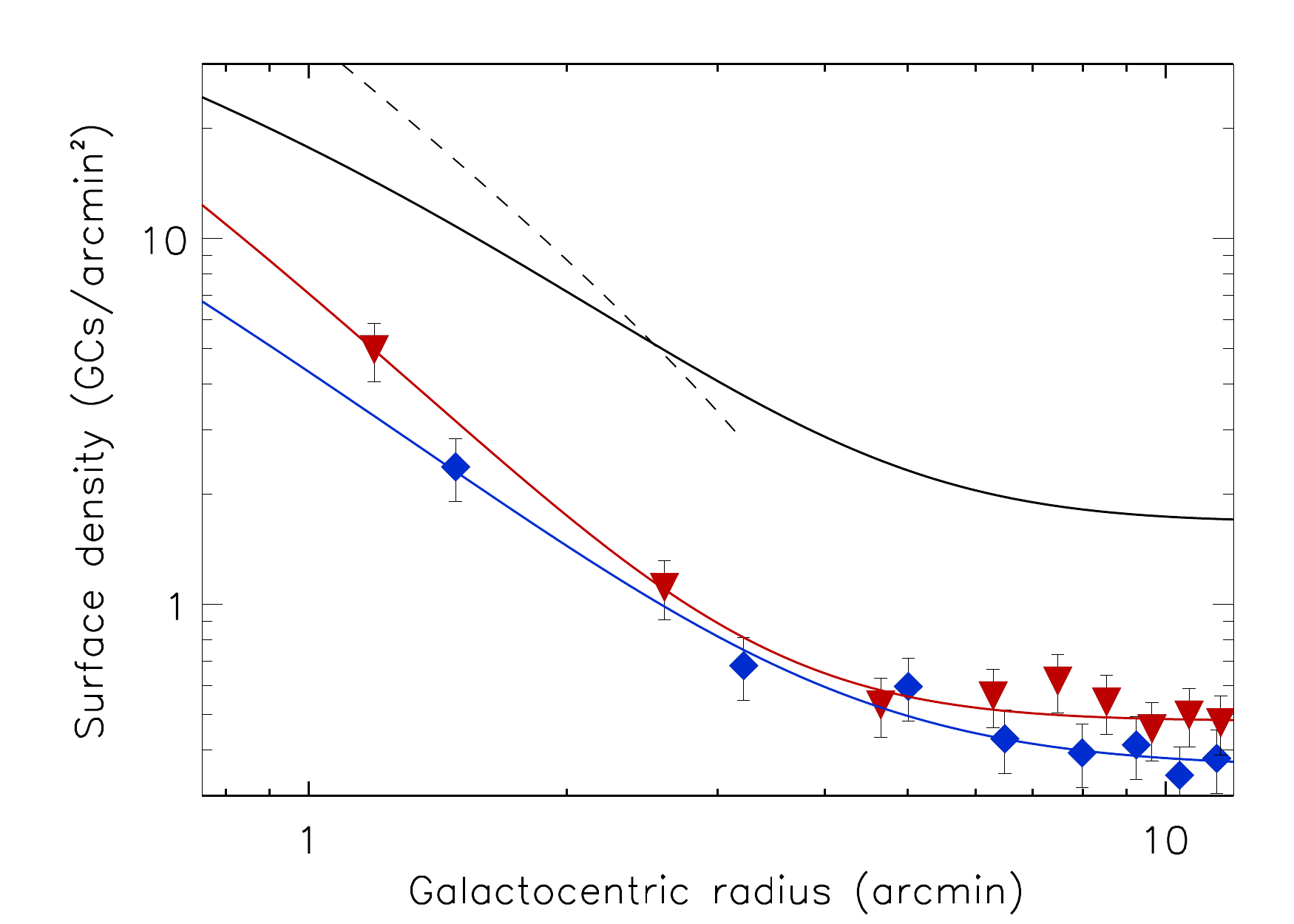} 
  \caption{Radial density distributions of GC subpopulations for NGC 3607. The density distributions for the blue and the red subpopulations  (from the MA method) are represented with blue diamonds and red triangles, respectively. The best fit S\'{e}rsic profiles to the density distributions are shown as solid lines. The black solid line represents the best fit S\'{e}rsic profile for the total GC system. The dashed line represents the galaxy brightness profile in the {\it i} filter. The blue subpopulation is found to be more extended than the red subpopulation. However, the galaxy stellar light profile better matches with the density distribution of the red subpopulation than the blue subpopulation. } 
  \label{BR_07}
\end{figure}

 For NGC 3607, the GMM algorithm confirmed a bimodal colour distribution from the SB and MA method selected GCs, based on the following statistics : with less than 0.001 percent confidence level, D > 2.6 $\pm$ 0.3 and negative kurtosis.  The blue and red GC subpopulations peak in ({\it g$-$i}) colour at 0.74 $\pm$ 0.04 and 1.03 $\pm$ 0.03, respectively.  The ({\it g$-$i}) colour of separation between the blue and the red subpopulations is at 0.87 $\pm$ 0.02. The total GC system is classified into 45 $\pm$ 9 and 55 $\pm$ 8 percent, respectively, blue and red subpopulations.

\begin{table}
\centering
\caption{Fitted parameters for the surface density profile of NGC 3607 and NGC 3608 GC subpopulations. The first and second columns represent the target galaxy and subpopulation category. The derived parameters, effective radius, the S\'{e}rsic index and the background estimation, after the S\'{e}rsic fit  are given in the last three columns.} 
\begin{tabular}{ccccc}
\hline
NGC&GC  & R$_e$ & n & bg \\ 
      &      & (arcmin) & & (arcmin$^{-2}$)  \\ 
\hline\hline
\multirow{2}{*}{3607} &Blue & 1.59 $\pm$ 0.94 & 4.14 $\pm$ 2.32  & 0.36 $\pm$ 0.12 \\
                               &Red & 0.67 $\pm$ 0.52 & 3.38 $\pm$ 1.35 & 0.48 $\pm$ 0.05 \\ 
\hline
\multirow{2}{*}{3608} &Blue & 1.42 $\pm$ 0.31 & 1.03 $\pm$ 0.89  & 0.50 $\pm$ 0.05 \\
                               &Red & 0.91 $\pm$ 0.72 & 1.98 $\pm$ 0.82 & 0.35 $\pm$ 0.05 \\ 
\hline
\end{tabular}
\label{surfden078}
\end{table}

\begin{figure*}
\centering
     \subfloat{\includegraphics[width=0.45\textwidth]{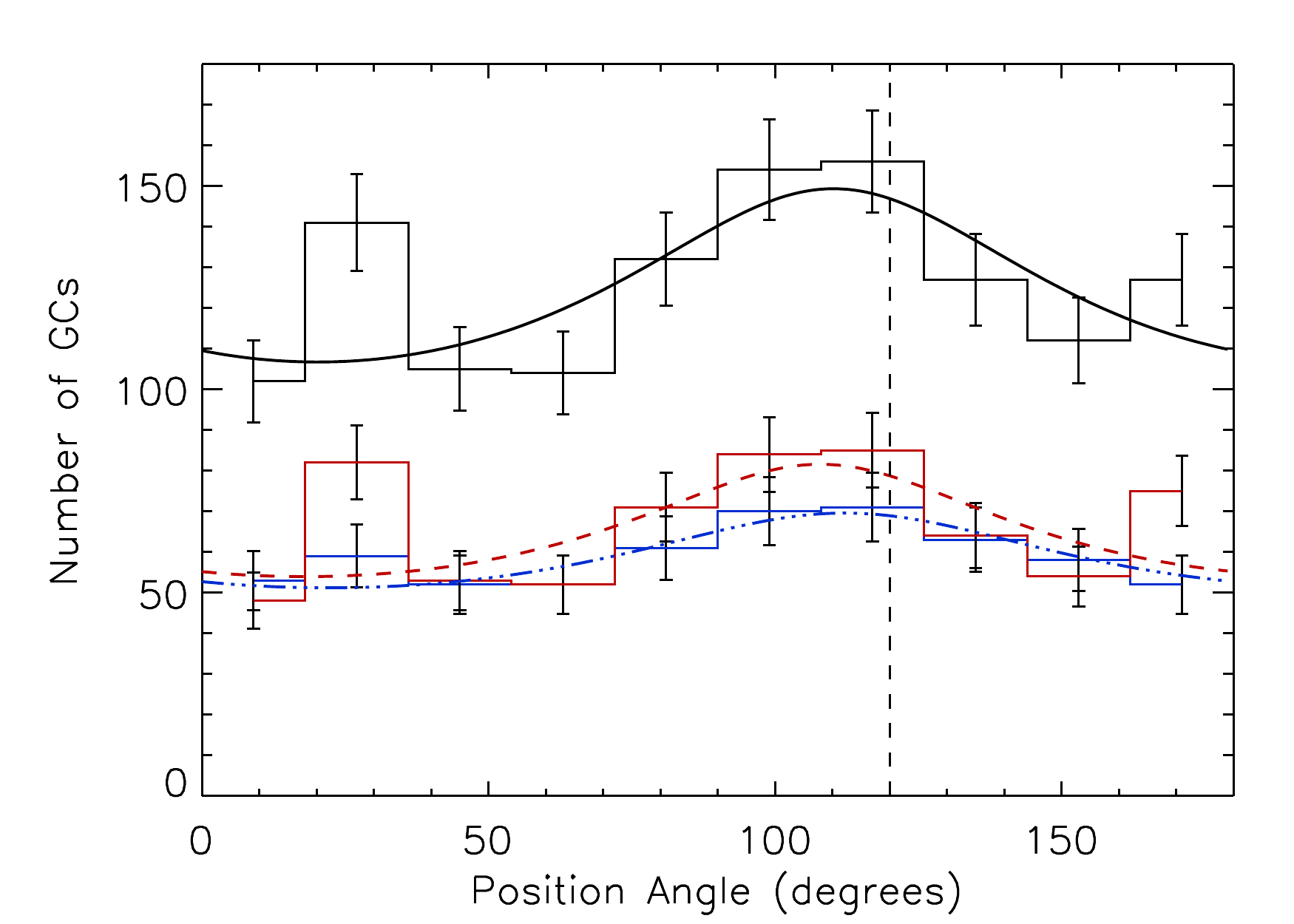}\label{AD_07_SB}}
     \subfloat{\includegraphics[width=0.45\textwidth]{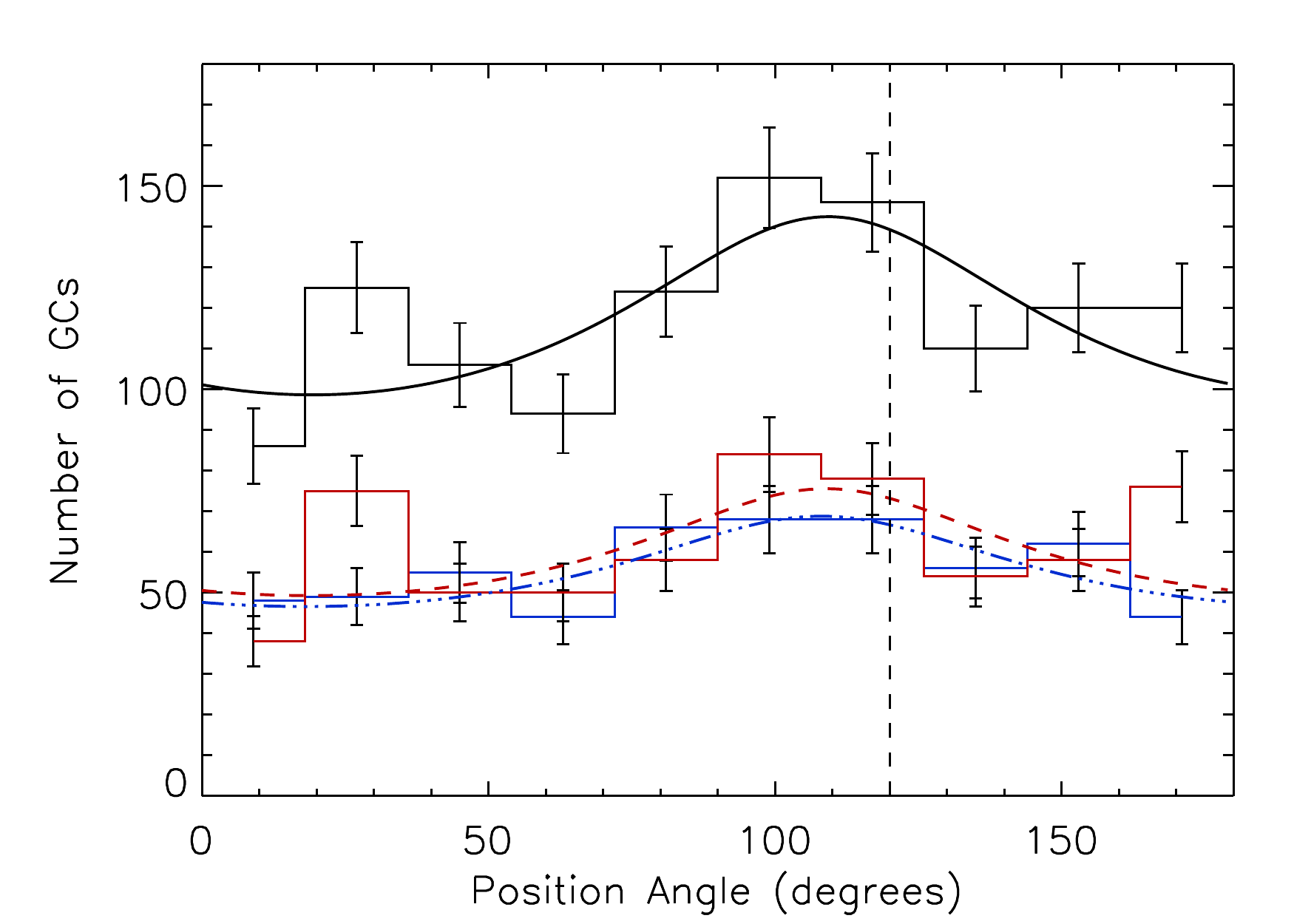}\label{AD_07_MA}}
\caption{Azimuthal distribution for the GC system of NGC 3607. The black, blue and red histograms represent the azimuthal distribution of the total population and the blue and red subpopulations of GCs. The solid, dashed and dash-dotted lines represent the fitted profiles based on Equation \ref{azimuth}. The vertical dashed line shows the position angle of the stellar major axis, 125 degrees. The {\it left} panel includes GCs selected on the basis of the SB method, whereas the {\it right} panel includes GCs based on the MA method. In both panels, the total GC system and the blue and red subpopulations are aligned in a position angle which is in good agreement with the stellar light. An overabundance of GCs (majority from the red subpopulation) along the minor axis (35 degrees) is seen in both panels. } 
\label{AD_07}
\end{figure*}

The radial density distribution for both GC subpopulations (from the MA method) are estimated and plotted in Figure \ref{BR_07}. Both subpopulation distributions are fitted with S\'{e}rsic profile given in Equation \ref{sersic}. The parameters derived from the S\'{e}rsic fit are tabulated in Table \ref{surfden078}. The red subpopulation is centrally concentrated while the blue subpopulation is more extended. The red subpopulation appears to have higher number density for most galactocentric radii. The galaxy stellar light profile is in better agreement with the density distribution of red subpopulation than blue subpopulation. Also the effective radius of the galaxy stellar light (39 arcsec) matches more with the red subpopulation (40 $\pm$ 29 arcsec) than the blue ones (95 $\pm$ 50 arcsec).

\subsubsection{Azimuthal distribution}
To quantify the azimuthal distribution of GCs, they are initially folded along the North to South direction, then binned in equal angular intervals.  The azimuthal distribution,  $\sigma(R,\theta)$, is then fitted with a profile \citep{McLaughlin1994} of the form: 
\begin{eqnarray}\nonumber
 \sigma(R,\theta) &=& kR^{-\alpha} \left[cos^2(\theta - \text{PA}) + \right. \\
                               && \left. (1-\epsilon^2)^{-2} sin^2(\theta - \text{PA})\right]^{-\alpha/2} + bg
\label{azimuth}
\end{eqnarray}
where {\it $\alpha$ }is the power law index fitted to the surface density of GCs, {\it bg} is the background estimated from the S\'{e}rsic fits (see  Section \ref{SD}) and {\it k} is the normalization constant. The profile is iterated with the position angle of the GC system (PA) and the ellipticity ($\epsilon$) as free parameters. For the analysis, only the GCs within the extent of GC system (i.e., 9.5 arcmin) are included. The number of GCs in each angular bin is corrected for the missing area due to NGC 3608 in the SB method, and  is doubled in the MA method.  Here we used 980 and 564 GCs, respectively, in the SB and MA methods.

Figure \ref{AD_07_SB} shows the azimuthal distribution of GCs selected based on the SB method. The GCs are aligned to a position angle of 110 $\pm$ 7 degrees, which is in reasonable agreement with the stellar light (125 degrees) of the galaxy. The alignment of GC system is more elliptical (0.39 $\pm$ 0.08) than the stars (0.13). The GCs also show an enhancement along the minor axis (35 degrees), which is either a genuine feature or possibly a contamination from the GCs of NGC 3608 and NGC 3605 (both positioned around the minor axis of NGC 3607).  We already found a constant surface density around NGC 3605 and hence, we assume that NGC 3605 is not contributing to the overabundance. 

The only other possible contributor for this minor axis overabundance is NGC 3608, situated in the NE direction. We have eliminated the maximum contamination from NGC 3608 in the MA method, as it counts only the hemisphere away from the other galaxy. Hence, if the enhancement of GCs is not genuine, then we should not observe the same in the MA method. Figure \ref{AD_07_MA} displays the azimuthal distribution of GCs selected in the MA method, including only the GCs from 125 to 305 degrees counted from North in counter-clockwise direction. It is evident from this plot that the enhancement along the minor axis is a genuine feature, with decreased strength which is consistent within error bars. The position angle of GCs from the MA method also aligns with the galaxy stellar light. Similarly, from the SB method, the GCs are found to be more elongated than the arrangement of stellar light. Table \ref{AD_07_08} summarises the best fit sinusoidal profile parameters.

\begin{table}
\caption{Position angle and ellipticity for the GC systems of NGC 3607 and NGC 3608. The values are derived by fitting Equation \ref{azimuth} to the azimuthal distribution. The table gives the derived values for the total GC system, the blue and the red subpopulations. For comparison, the position angle and the ellipticity of the galaxy stellar light for NGC 3607 are 125 degrees and 0.13, respectively and for NGC 3608 are 82 degrees and 0.20, respectively. }
\begin{tabular}{cccc|cc} 
\hline\hline
Method & \multicolumn{3}{c}{NGC 3607} & \multicolumn{2}{c}{NGC 3608} \\
\hline
                     &GC&PA&{\large$\epsilon$}&PA&{\large$\epsilon$} \\
                     &      &($^o$)&         & ($^o$) & \\
\hline               
\multirow{3}{*}{SB} &Total     &110 $\pm$ ~7      & 0.39 $\pm$ 0.09          & 104 $\pm$ 15   & 0.20 $\pm$ 0.09   \\
                              &Blue      & 112 $\pm$ 14   &   0.37 $\pm$ 0.11    &  106 $\pm$ 11  &  0.31 $\pm$ 0.10  \\
                             & Red      & 108 $\pm$ 11   &  0.47 $\pm$ 0.09         & ~97 $\pm$ 18   & 0.14 $\pm$ 0.16   \\
    \hline                         
\multirow{3}{*}{MA}&Total     &109 $\pm$ ~8     &  0.42 $\pm$ 0.07         & ~66 $\pm$ ~7   &  0.39 $\pm$ 0.10  \\
                              & Blue    & 108 $\pm$ 10   &  0.45 $\pm$ 0.11        & ~67 $\pm$ ~8    & 0.45 $\pm$ 0.09   \\
                              & Red    & 109 $\pm$ ~8     & 0.48 $\pm$ 0.11            &  ~64 $\pm$ 10  & 0.44 $\pm$ 0.13   \\

\hline
\end{tabular}
\label{AD_07_08}
\end{table}

Figure  \ref{AD_07} also shows the azimuthal distribution of blue and red GC subpopulations from the two methods. The subpopulations are separated at a ({\it g$-$i}) colour of 0.87, obtained from the GMM algorithm. Both subpopulations have similar position angles for the total GC system and are more elliptical than the galaxy stars. 

Summarising, the total GC system and both subpopulations follow the galaxy stellar light in position angle. But the distribution of GCs is not as circular as the galaxy stellar component. The red GC subpopulation shows a more flattened distribution than the blue subpopulation for NGC 3607.

\begin{figure}
\centering
 \includegraphics[scale=0.5]{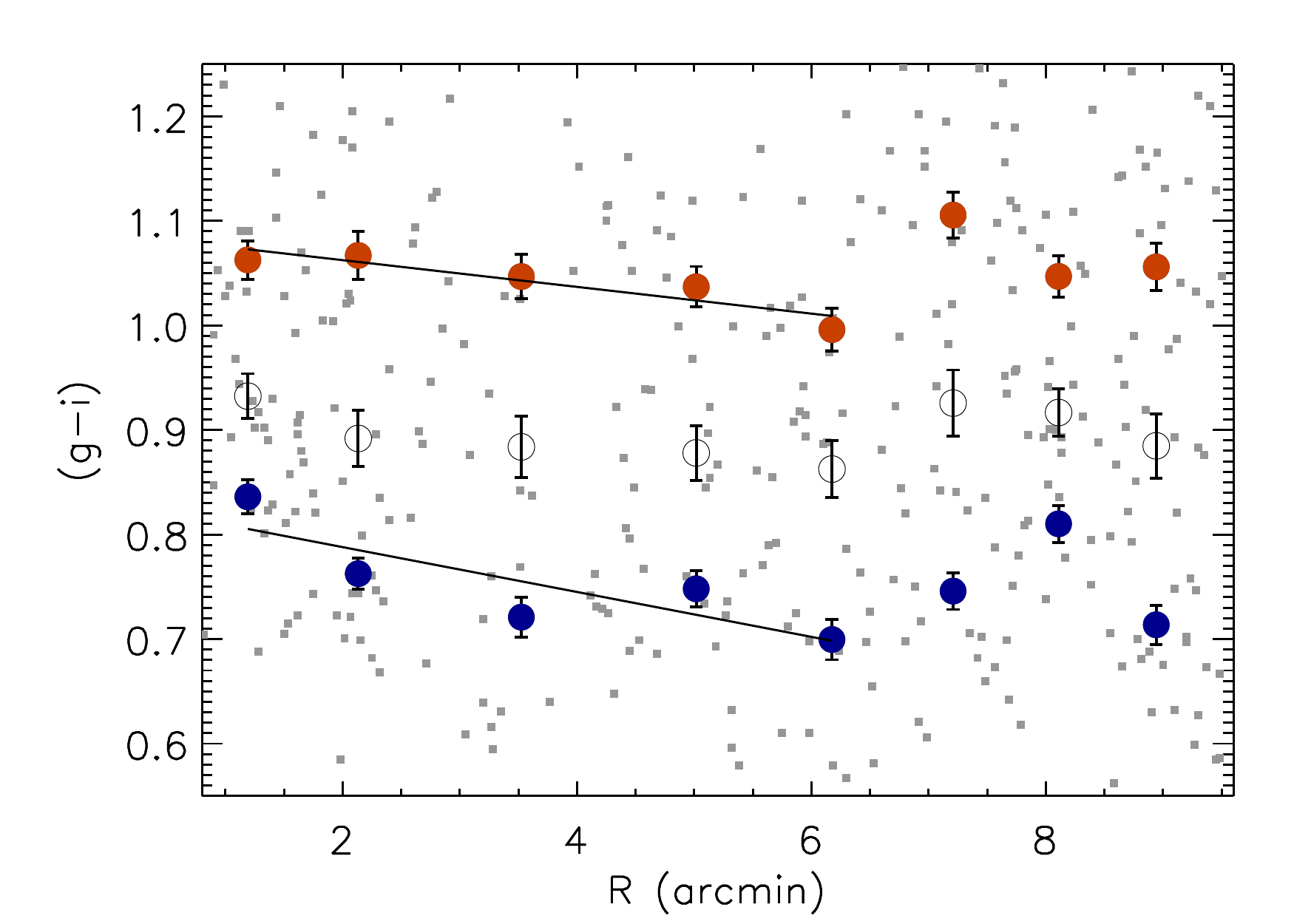}
  \caption{Radial colour distribution for the GC system of NGC 3607.  The GCs are selected using the MA method, and are shown as small grey squares.  The separation between the two subpopulations is obtained using a moving mean colour, and shown in black open circles. The average colours with errors for the blue and the red GC subpopulations are shown as blue and red filled circles, respectively. The solid lines represent the best fit lines for the blue and the red subpopulations in the  central 6.5 arcmin, the projected separation between the two galaxies. For the blue and the red GC subpopulations, significant colour gradients ($-$0.070 $\pm$ 0.013 and $-$0.033 $\pm$ 0.015 mag per dex for blue and red GC, respectively) are obtained in the central 6.5 arcmin radius. }
\label{CD_07}
\end{figure}

\subsubsection{Radial colour distribution}
Figure \ref{CD_07} shows the radial distribution of GC colours from the centre of NGC 3607. The GCs brighter than the turnover magnitude in the MA method only are included. The GC subpopulations are divided with a moving colour with radius technique. In each radial bin, the average colour for both subpopulations are determined (keeping  a constant number of GCs per radial bin). For NGC 3607, we used $\sim$ 350 GCs to plot the colour distribution with 45 GCs in each bin.

As seen from the plot, for the total extent of the GC system, the average colour for the blue subpopulation decreases with radius from the centre, while a flat colour gradient is seen for the red subpopulation. The colour distribution for the blue subpopulation is fitted with a logarithmic relation (following \citealt{Forbes2011}) as:
\begin{equation}
(g-i) = a + b\times\text{log}(R/R_e)
\label{colourdis}
\end{equation}
where R$_e$ is the effective radius for NGC 3607 equal to 39 arcsec \citep{Brodie2014}, {\it a} and {\it b} are, respectively, intercept and slope of the fit. We obtained a best fit line using the bootstrap technique and derived the parameters for the blue subpopulation as {\it a} = 0.82 $\pm$ 0.018 and {\it b} = $-$0.036  $\pm$ 0.009 mag per dex. \citet{Maraston2005} derived a relation between ({\it g$-$i}) and [{\it Z/H}] over the metallicity range [{\it Z/H}] $\le$ $-$0.2, using single stellar population models, of $\Delta$({\it g$-$i})/$\Delta$[{\it Z/H}] = 0.21 $\pm$ 0.05 mag per dex. Using this we obtained for the blue subpopulation a metallicity gradient of $-$0.17 $\pm$ 0.04 dex per dex to the total extent of the GC system. But, we did not detect a significant colour gradient for the red subpopulation and the total population in the total extent of GC system ($-$0.01  $\pm$ 0.01 and $-$0.013  $\pm$ 0.011 mag per dex for red and total GCs). 

 We also quantified the colour/metallicity gradient in the central ($\sim$ 6.5 arcmin) region, only including the common galactocentric radii between the two galaxies. The colour gradient for the blue, red and the total population are  $-$0.070 $\pm$ 0.013, $-$0.033 $\pm$ 0.015 and $-$0.039 $\pm$ 0.018 mag per dex. In the inner 6.5 arcmin region, the blue subpopulation has a higher metallicity gradient ($-$0.33 $\pm$ 0.06 dex per dex) compared to the red subpopulation ($-$0.16 $\pm$ 0.07 dex per dex). Hence, we conclude that a significant colour/metallicity gradient is obtained for the blue and the red subpopulations of NGC 3607. 

\begin{figure}
\centering
 \includegraphics[scale=0.5]{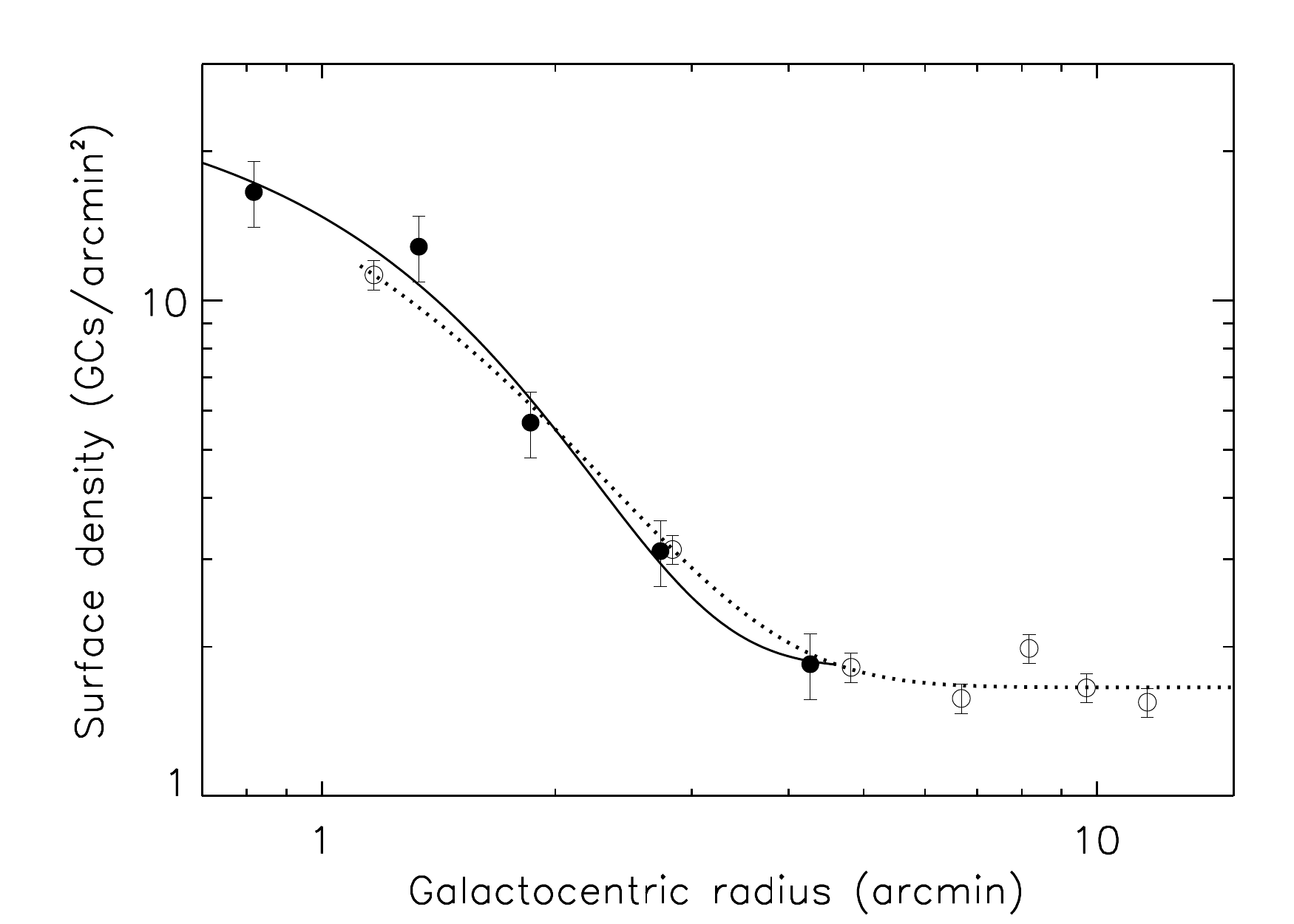} 
  \caption{Surface density distribution for the GC system of NGC 3608. The radial density distribution of GCs based on the SB method and the MA method are represented with filled and open circles, respectively. The SB method detects GCs to a maximum galactocentric radius of $\sim$ 5.5 arcmin. The best fit S\'{e}rsic profiles are represented with solid and dotted lines for the two methods. The GC system reaches a background in the MA method at a galactocentric radius of 6.6 $\pm$ 0.8 arcmin. } 
  \label{SD_08}
\end{figure}

\subsection{GC system of NGC 3608}
\subsubsection{Radial density distribution}
Figure \ref{SD_08} displays the radial density of GCs selected with the SB and the MA methods for NGC 3608 fitted with the profile given in Equation \ref{sersic} (fitted parameters are given in Table \ref{surfden08}). In the SB method, the selection of GCs for NGC 3608 gives a maximum galactocentric radius of $\sim$ 5.5 arcmin (as seen from Section \ref{SB}). But the MA method identifies objects to a distance of 12.8 arcmin from the galaxy centre (thus extends up to the edge of the detection area). In both methods, the GCs with {\it i} $<$ 23.5 mag (turnover magnitude) are counted for studying this distribution.  In the density distribution plot, the SB and MA methods used 304 and 402 objects, respectively. The density distribution of GCs in radial annuli, after applying respective corrections for both methods, are shown in Figure \ref{SD_08}. The GC system reaches a background level of 1.65 $\pm$ 0.1 GCs per square arcmin to a galactocentric radius of 6.6 $\pm$ 0.8 arcmin (43 $\pm$ 5 kpc), from the MA method. But the density value for the final data point from the SB method is 1.82 $\pm$ 0.36 GCs per square arcmin implying that the distribution has not reached the background level.  The elimination of marginal GCs (SB probability between 50 and 55 percent) in the SB method might be the reason for this discrepancy in the extent of GC system. Another point from the figure is that the surface density values estimated from both methods are consistent within error bars, up to 5.5 arcmin.

\begin{table}
\centering
\caption{Fitted parameters for the surface density of NGC 3608 GC system. The first column represents the GC selection method. The following three columns give the derived values for the effective radius, the S\'{e}rsic index and background using the S\'{e}rsic fit. The extent of the GC system is given in the last column, which is not estimated for the SB method.}
\begin{tabular}{ccccc}
\hline
Method  & R$_e$ & n & bg  & GCS ext.\\ 
            & (arcmin) & & (arcmin$^{-2}$) & (arcmin) \\ 
\hline\hline
SB & 1.29 $\pm$ 0.15 &0.66 $\pm$ 0.36 & 1.82 $\pm$ 0.36 &  -\\
MA & 1.50 $\pm$ 0.15 &0.93 $\pm$ 0.56 &1.65 $\pm$ 0.10 &6.6 $\pm$ 0.8\\ 
\hline
\end{tabular}
\label{surfden08}
\end{table}

\begin{figure}
\centering
 \includegraphics[scale=0.33]{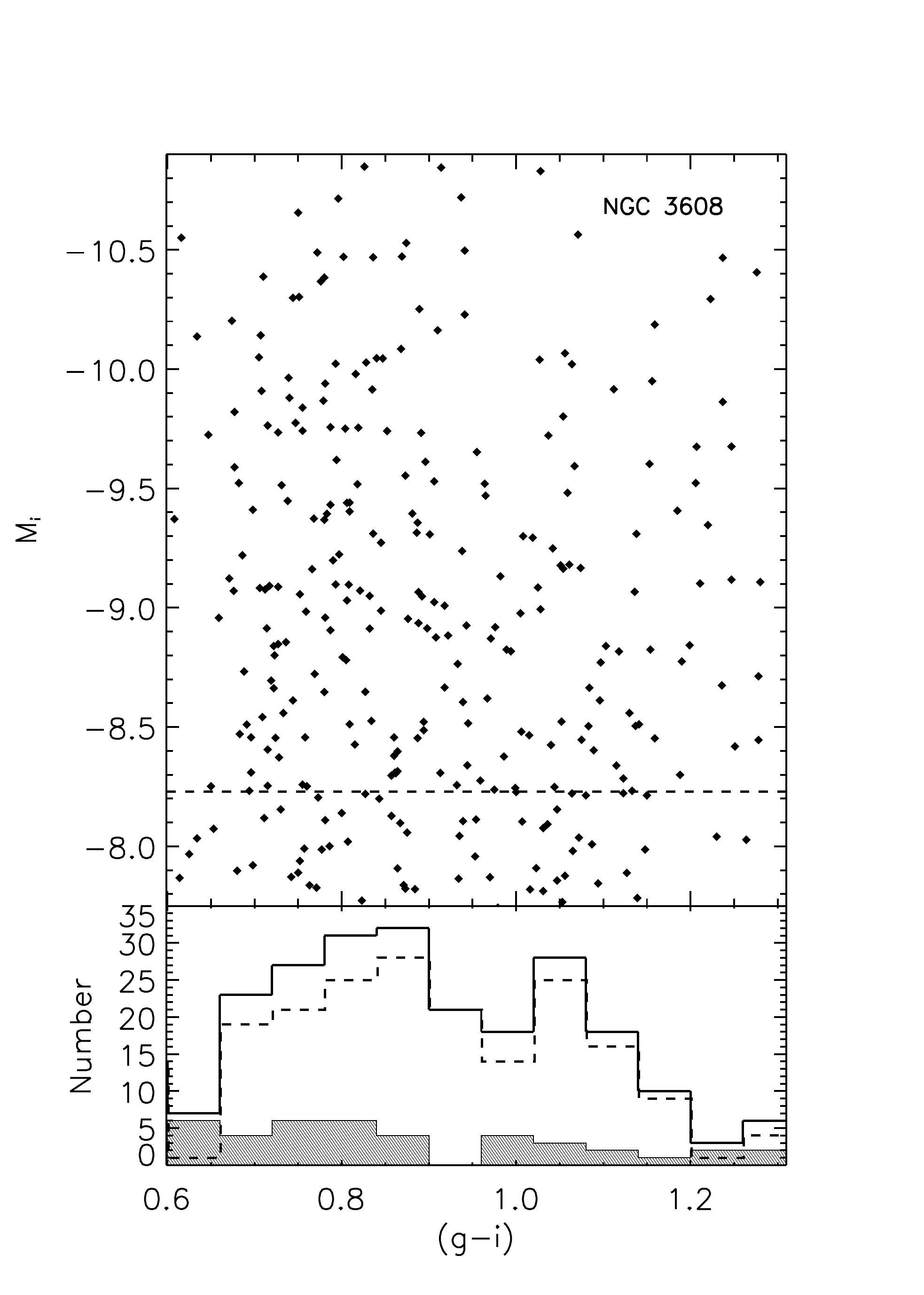}
   \caption{ Colour magnitude diagram for NGC 3608. The top panel represents the GCs brighter than M$_i$ = $-$7.75 mag within the extent of GC system. The dashed line represents the turnover magnitude in {\it i} filter, M$_i$ = $-$8.23 mag. The colour histogram of the GC system of NGC 3608 is shown in the bottom panel, where the open, shaded and dashed histograms represent the GCs which are brighter than the turnover magnitude, the estimated background contamination and the background corrected colour histograms. } 
  \label{CMD_08}
\end{figure}

NGC 3608 is an E2 galaxy and M$_V^T$ = $-$20.98 mag, assuming a  mass to light ratio of 10 \citep{Zepf1993} has a stellar mass of log (M/M$_{\odot}$) = 11.32. Using Equation \ref{GCSextent}, the expected GC system extent is calculated to be 40 $\pm$ 2 kpc, consistent with the GC system extent from the observational data (43 $\pm$ 5 kpc).

\subsubsection{GC bimodality}
\label{8subpop}
 The colour magnitude diagram for the selected GCs of NGC 3608, within the GC system extent (43 kpc) and brighter than M$_i$ = $-$7.75 mag, is shown in Figure \ref{CMD_08}. The figure displays $\sim$ 250 GCs.  The background contamination in the GC system selection is quantified, as explained in Section \ref{GCb}, and are corrected for this contamination. The bottom right panel displays the colour histograms of GCs which are brighter than the turnover magnitude with and without background correction. The estimated background correction is also illustrated in the same figure.

The GMM algorithm fit to NGC 3608 GCs selected from the MA method gives a bimodal colour distribution with peaks at ({\it g$-$i}) = 0.80 $\pm$ 0.02 and 1.12 $\pm$ 0.04. The total GC system contains 65 $\pm$ 6 and 35 $\pm$ 6 percent, respectively, blue and red subpopulations. The blue and red subpopulations are divided at ({\it g$-$i}) = 0.93.

\begin{figure}
\centering
 \includegraphics[scale=0.5]{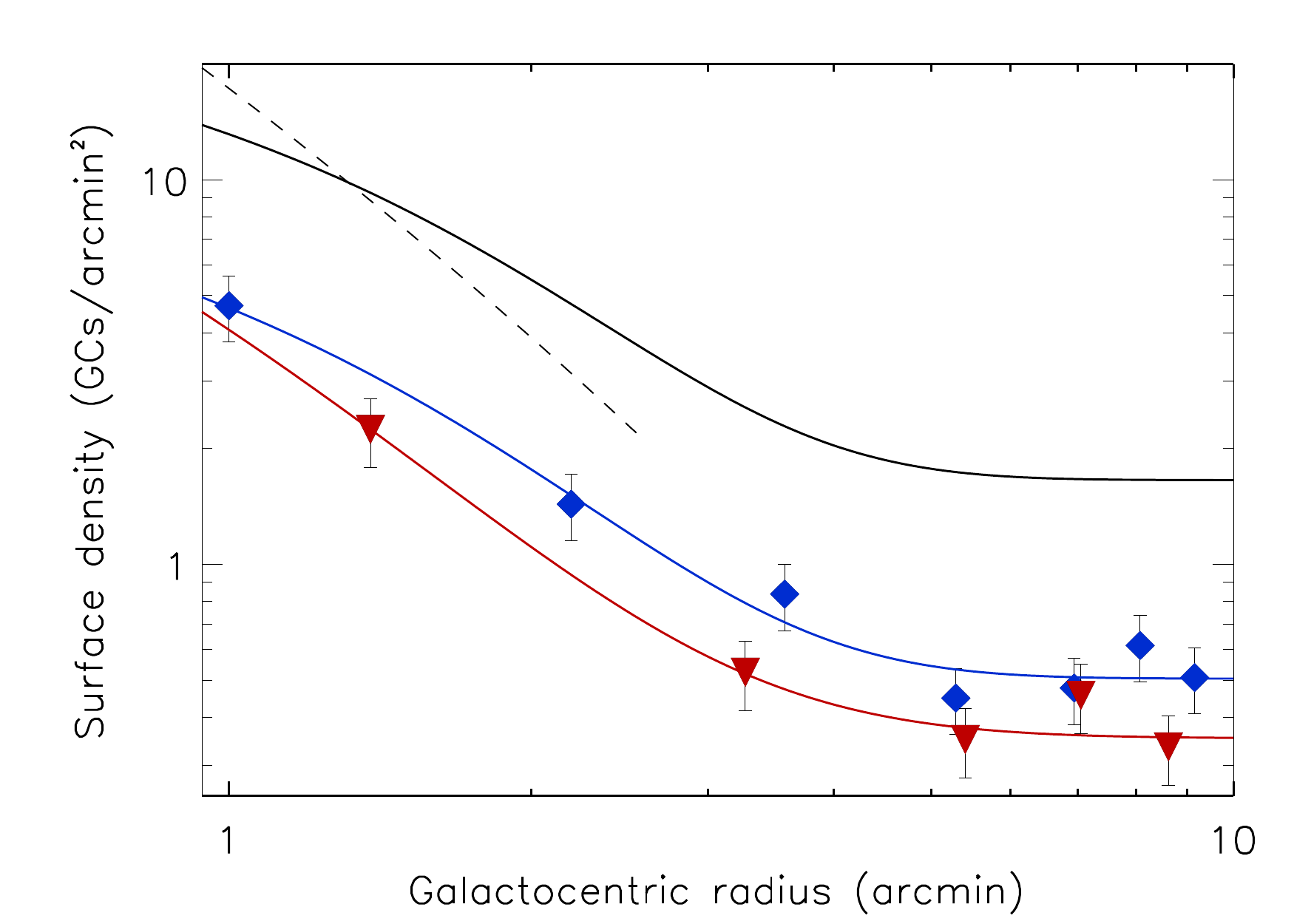} 
  \caption{ Radial density distributions of NGC 3608 GC subpopulations. The blue diamonds and the red triangles represent the surface density distributions of blue and red subpopulations respectively. The blue and the red solid lines demonstrate the best fit S\'{e}rsic profiles on the distributions, while the black solid line represents the S\'{e}rsic fit for the total GC system. The galaxy brightness profile in the {\it i} filter is shown as dashed line, in reasonable agreement with the density distribution of red subpopulation. Also, the red subpopulation is more centrally concentrated than the blue subpopulation for NGC 3608. } 
  \label{BR_08}
\end{figure}

The radial surface densities (GCs from the MA method) are fitted with S\'{e}rsic profiles and are displayed in Figure \ref{BR_08}. The parameters estimated from the S\'{e}rsic fit are tabulated in Table \ref{surfden078}. For NGC 3608, the blue subpopulation shows a higher density than the red subpopulation throughout the extent of the GC system. The red subpopulation is found to be more centrally concentrated, and their density profile is in good agreement with the galaxy stellar light. 

\begin{figure*}
\centering
     \subfloat{\includegraphics[width=0.45\textwidth]{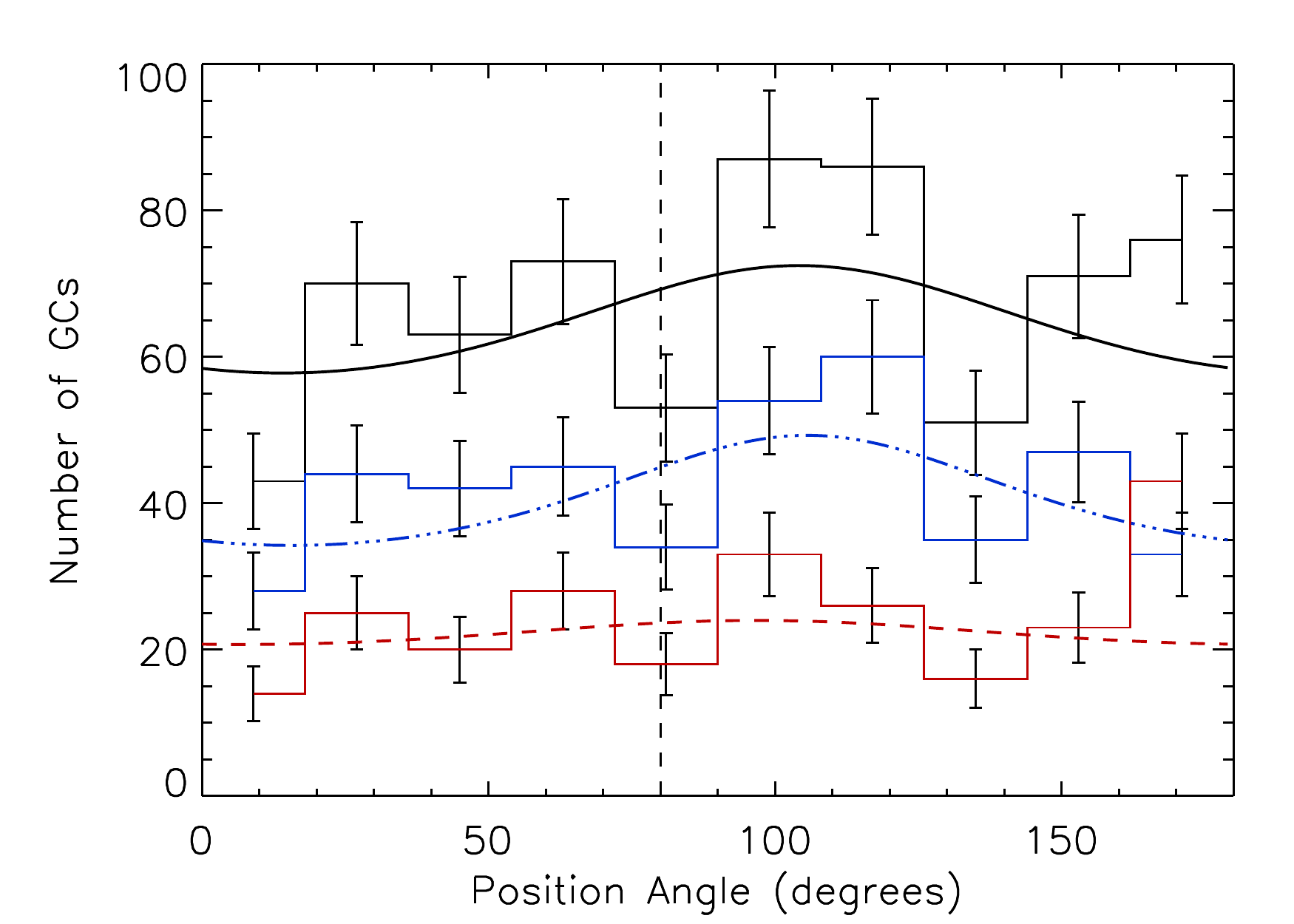}\label{AD_08_SB}}
     \subfloat{\includegraphics[width=0.45\textwidth]{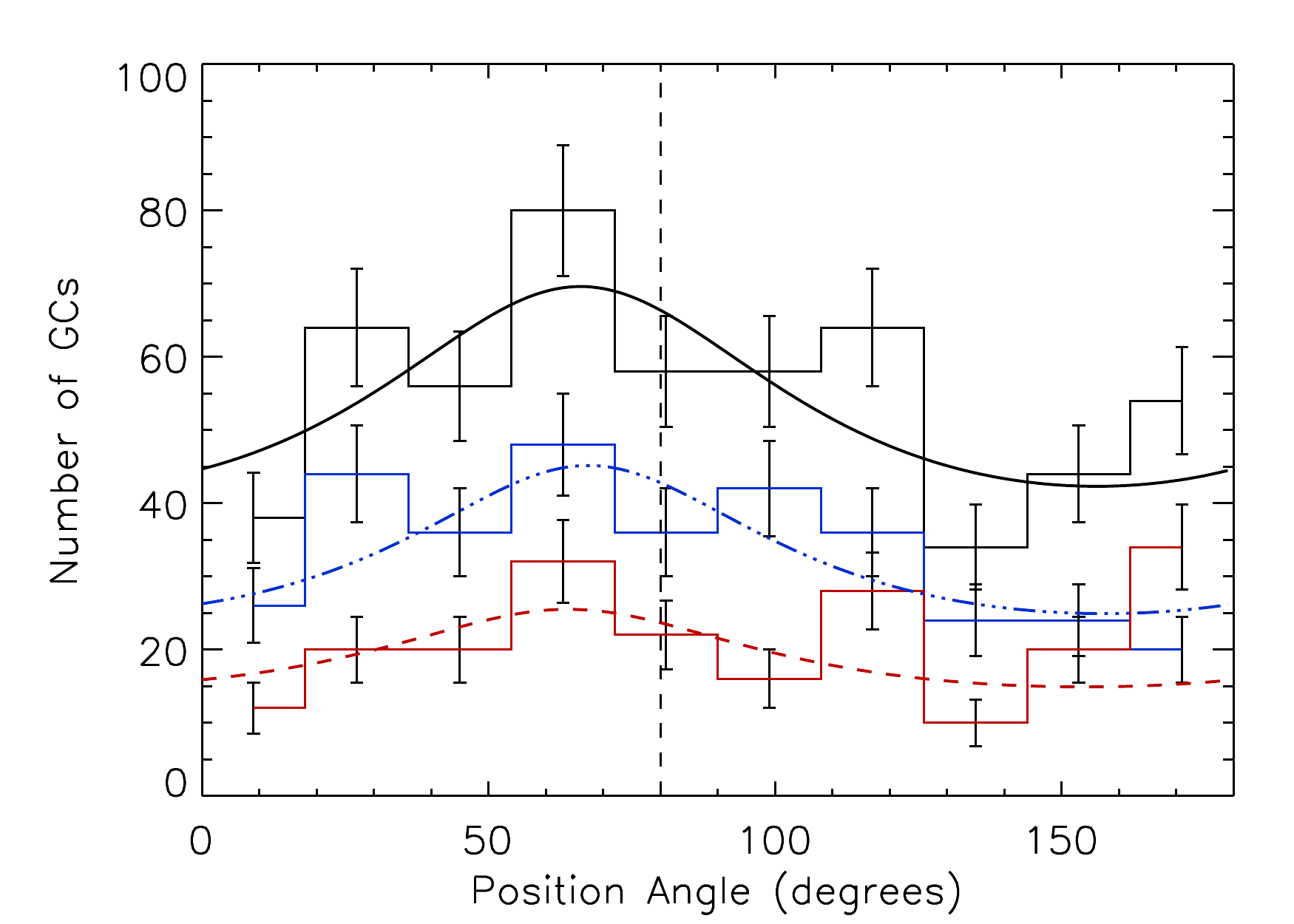}\label{AD_08_MA}}
\caption{Azimuthal distribution for the GC system of NGC 3608. The colours and styles of histograms and lines are same as shown in Figure \ref{AD_07}. The {\it left} panel shows the distribution of GCs selected from the SB method and the right panel shows the distribution from the MA method. The galaxy stellar light is aligned along the major axis (82 degrees), represented by the vertical dashed line. The total GC system and both the subpopulations are arranged along a different position angle ($\sim$ 100 degrees) than the galaxy stellar light in the SB method. The {\it right} panel shows the distribution of GCs selected from the MA method. The total GC system and both the subpopulations are aligned at a position angle, $\sim$ 65 degrees in the MA method, slightly off from the galaxy stellar light. Also in both panels, a scarcity of GCs along the galaxy major axis is visible.  } 
\label{AD_08}
\end{figure*}

\subsubsection{Azimuthal distribution}
The range of galactocentric radii for the selected GCs in the SB method is from 0.5 to 5.5 arcmin. The selection of GCs  in all position angles  is complete up to 2.2 arcmin and hence, an areal correction is applied for the missing area outside that radius. Here we used 378 and 275 GCs, respectively, in the SB and MA methods. Figure \ref{AD_08_SB} shows the azimuthal density distribution of GCs from the SB method. The histograms are fitted with the sinusoidal profile given in Equation \ref{azimuth}.  Table \ref{AD_07_08} gives the position angles and ellipticities obtained from the sinusoidal fit. The galaxy stellar light has a major axis of 82 degrees and ellipticity of 0.20.  As seen from Table \ref{AD_07_08}, the total GC system and both subpopulations are arranged along a different position angle of $\sim$ 100 degrees for the SB method. When the distribution is examined over 0 to 360 degrees rather than 0 to 180 degrees (i.e., without folding along the North to South direction), an overabundance  is evident in the position angles between 90 and 230 degrees. This is in the direction towards NGC 3607 and also the direction in which the area correction is largest. Hence, this overabundance is either due to contamination from NGC 3607 (or due to overestimation of missing area). Also a scarcity of GCs is observed in both major axis position angles (82 and 262 degrees). The ellipticity for the total GC system is 0.20 $\pm$ 0.09, matching with the galaxy stellar light.

Figure \ref{AD_08_MA} shows the azimuthal density distribution of GCs selected in the MA method, for which GCs in the position angles 80 to 260 degrees are under abundant. The GCs within the extent of GC system (6.6 arcmin) are included in the azimuthal distribution. As seen in Table \ref{AD_07_08}, the best fit sinusoidal profile gives a position angle of 66 $\pm$ 7 degrees for the total GC system and an ellipticity of 0.39 $\pm$ 0.10.  The GCs selected in the MA method includes GCs of NGC 3608 placed at a position angle pointing away from NGC 3607, implying minimum contamination. The arrangement of GCs in the MA method is along the position angle matching the galaxy stars, but the distribution is more elliptical.  Since we observed an overabundance in GCs for both galaxies, in the region towards each other, an interaction may be occurring between the two. 

\begin{figure}
\centering
 \includegraphics[scale=0.5]{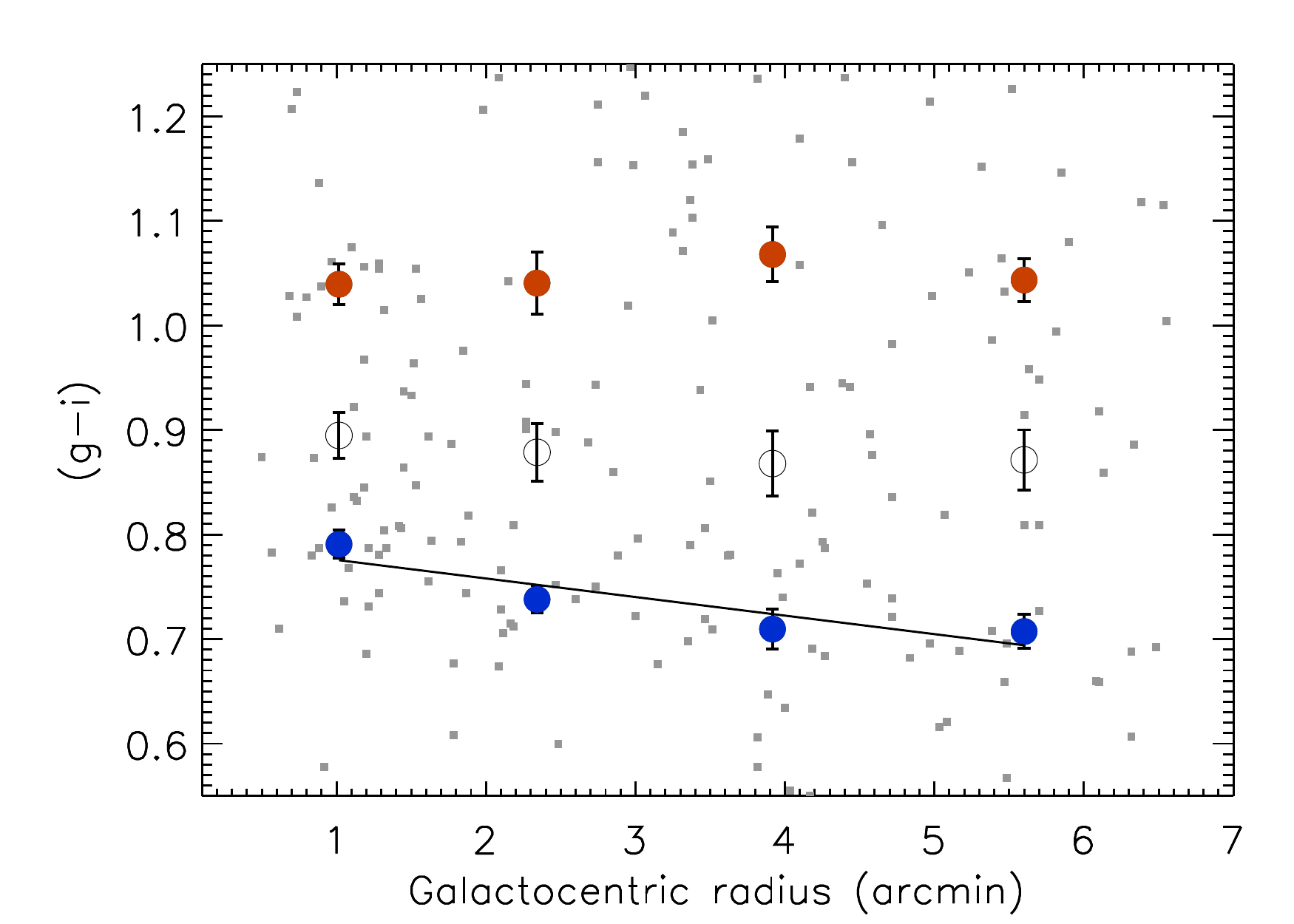} 
  \caption{Radial colour distribution for the GC system of NGC 3608 (from the MA method). The individual GCs are represented as small grey squares. The average colours with error for the blue and red GC subpopulations are denoted as blue and red filled circles, respectively. The separation colours for the subpopulations in each radial bin is calculated using a moving mean colour method and is denoted with black open circles. The blue subpopulation shows a colour gradient of $-$0.052 $\pm$ 0.011 mag per dex ($\Delta$[Z/H] = $-$0.25 $\pm$ 0.05 dex per dex) for the total extent of the GC system, but we did not detect any colour gradient for the red subpopulation.} 
\label{CD_08}
\end{figure}

The total GC system is separated into subpopulations at ({\it g$-$i}) = 0.93 (obtained from the GMM algorithm). Regarding the azimuthal distribution of GC subpopulations, both subpopulations are aligned along the position angle of the total GC system in the two methods. Also the ellipticity of both subpopulations matches with the total GC system. In the MA method, the total and both subpopulations are more elliptically aligned than the galaxy stellar light.

\subsubsection{Radial colour distribution}
Figure \ref{CD_08} shows the radial ({\it g$-$i}) colour distribution of GC system of NGC 3608 selected on the MA method. To study this distribution, GCs (from the MA method) brighter than the turnover magnitude are selected and is $\sim$ 215 GCs.  The total GC system and the red subpopulation show a null gradient, while the blue subpopulation shows a strong gradient along the total radial extent of the GC system. The colour distribution of the blue subpopulation is fitted with the logarithmic relation given in Equation \ref{colourdis}, where R$_e$ = 30 arcsec \citep{Brodie2014}. The parameters, {\it a} = 0.823 $\pm$ 0.019 mag and  {\it b} = $-$0.052 $\pm$ 0.011 mag per dex, are derived from the best fit profile using the bootstrap technique (shown in  Figure \ref{CD_08}). The colour gradient, when converted  to a metallicity gradient, gives $\Delta$[{\it Z/H}] = $-$0.25 $\pm$ 0.05 dex per dex.

\section{Discussion}
\subsection{GC system distribution and galaxy effective radius}
In this study of two group galaxies (NGC 3607 and NGC 3608 situated within a projected distance of 39 kpc), we introduce two methods, the Surface Brightness and the Major Axis methods,  to separate the individual GC systems. For NGC 3607, the radial GC system extent determined from both methods are consistent with each other and in good agreement with the empirical relation for GC system extent (Equation \ref{GCSextent}), initially presented in \citet{Kartha2014}. From the radial surface density distribution, the red subpopulation is more centrally concentrated than the blue subpopulation.  The galaxy surface brightness distribution is in agreement with the density distribution profile of the red subpopulation than the blue subpopulation (Figure \ref{BR_07}). Also, the effective radius of the galaxy stars (39 arcsec) is consistent with that  of the red GC subpopulation (40 $\pm$ 29 arcsec), while for the blue GC subpopulation it is 95 $\pm$ 50 arcsec. Both the spatial distribution and the effective radius measurements support the idea that the red GC subpopulation has evolutionary similarities with the galaxy stellar component \citep{Forbes2001a,Larsen2001,Brodie2006,Spitler2010,Forbes2012a}. 

For NGC 3608, the blue GC subpopulation is more extended than the red GC subpopulation. It is evident from Figure \ref{BR_08} that the density distribution of the red GC subpopulation  follows the galaxy stellar light distribution. However, the effective radius of galaxy light (30 arcsec) is half of the red subpopulation (59 $\pm$ 40 arcsec) and one third of the blue subpopulation (85 $\pm$ 18 arcsec). The effective radius of the red GC subpopulation is therefore not consistent with the stellar light component. Even so the resemblance of the density distribution profile with the galaxy stellar light might imply a significant association. 

\begin{figure}
\centering
 \includegraphics[scale=0.5]{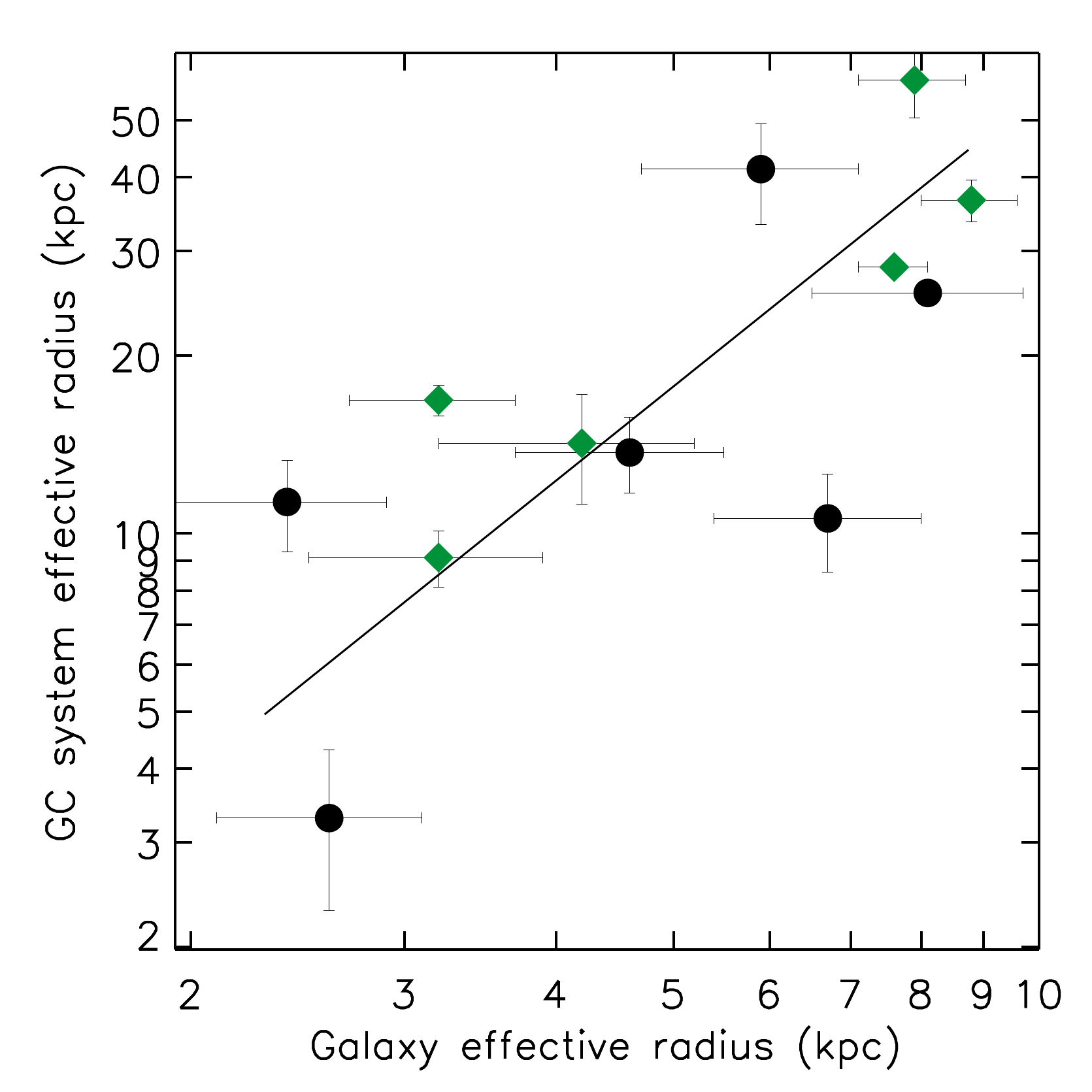} 
  \caption{GC system effective radius versus galaxy effective radius. The plot displays NGC 3607, NGC 3608, NGC 4406, NGC 4472, NGC 4594 and NGC 5813 (represented by green diamonds) along with other six galaxies (filled black circles) from the earlier study of \citet{Kartha2014}. The GC system effective radius is derived from the S\'{e}rsic profile fitted to the radial surface density distribution of GCs. The  data in the plot are fitted with a linear relation using the bootstrap technique shown by a black line. The GC system effective radius is $\sim$ 6 times the galaxy stellar light.} 
\label{Re_GCRe}
\end{figure}

\begin{table}
\centering
\caption{Effective radii for the GC systems  and respective stellar surface brightness, from the two galaxies in this paper (NGC 3607 and NGC 3608) and other four galaxies. The last column provides the references for the GC system and galaxy effective radii, respectively.}
\begin{tabular}{cccc}
\hline
\multicolumn{1}{c}{Galaxy}&\multicolumn{2}{c}{Effective radius}&\multicolumn{1}{c}{Ref.}\\
\hline
NGC   & GC system (kpc) & Stellar light (kpc)&\\ 
\hline\hline
3607   &   14.2$\pm$2   &   4.2$\pm$1 & 1, 2 \\
3608  &    9.1$\pm$1  &    3.2$\pm$0.7   & 1,  2\\
4406  &    28.2$\pm$1  &    7.6$\pm$0.5 & 3, 4 \\  
4472  &   58.4$\pm$8  &     7.9$\pm$0.8  & 3,  4\\ 
4594  &    16.8$\pm$1  &    3.2$\pm$0.7  & 3,  4\\ 
5813  &    36.6$\pm$3  &    8.8$\pm$0.8 & 3,  4\\ 
\hline
\end{tabular}
\newline
References: 1 - This work;  2 - \citet{Brodie2014}; 3 - \citet{Hargis2014}; 4 - \citet{Cappellari2011} 
\label{Re_table}
\end{table}

Figure \ref{Re_GCRe} displays the total GC system effective radius versus the galaxy effective radius and is an updated version (with the addition of six galaxies) of figure 20 in \citet{Kartha2014}. In this figure, the GC system effective radii are determined from S\'{e}rsic profile fits to the density distribution, which is currently available for twelve galaxies. The positions of the newly added galaxies, tabulated in Table \ref{Re_table}, are compatible with the existing linear relation (R$_{e(\text{GCS})}$ = [(5.2 $\pm$ 3.7) $\times$ R$_{e(\text{galaxy})}$] $-$ (8.5 $\pm$ 6.5)). The updated relation for the twelve galaxies is as follows:
\begin{equation}
R_{e(\text{GCS})} = [(6.5 \pm 1.3) \times R_{e(\text{galaxy})}] - (13 \pm 6) \\
\label{GCS}
 \end{equation}
 where both {\it R$_e$}s are measured in kpc. When compared with the relation in \citet{Kartha2014}, the above relation has a similar slope within error bars. The effective radii for both  GC subpopulations are determined only for six galaxies. With the available data, we could not detect any significant relation between the effective radius of GC subpopulations and the host galaxy stellar light. 

From Equation \ref{GCS}, we can infer that the GC system effective radius is $\sim$ 6 times the galaxy effective radius, which confirms that the GC system of a galaxy extends further out than the bulk of its stellar component \citep{Harris2000,Forbes2006,Brodie2006,Alamo2012,Cantiello2015}.  A byproduct from the above relation is that we can estimate the GC system effective radius by knowing the galaxy effective radius.

\subsection{GC system ellipticity and galaxy ellipticity}
To further address the association of galaxy stellar light with GC subpopulations, we need to study the two dimensional spatial distribution of these systems. Different studies of two dimensional distributions (position angle and ellipticity) have confirmed an association of both subpopulations with galaxy stellar light (e.g. NGC 2768 by \citealt{Kartha2014}, NGC 4636 by \citealt{Dirsch2005}). \citet{Park2013} analysed the two dimensional shape parameters of 23 early-type galaxies using the {\it HST}/ACSVCS. They found that the arrangement of both subpopulations is aligned with the photometric major axis of galaxies. Also, the red GC subpopulations show a tight relation in ellipticity with galaxy stellar light, while the blue GC subpopulations show a less tight relation. Concurrently, \citet{Kartha2014} obtained a similar relation for the red subpopulations from a sample of six galaxies using wide-field imaging. 

\begin{figure}
\centering
 \includegraphics[scale=0.5]{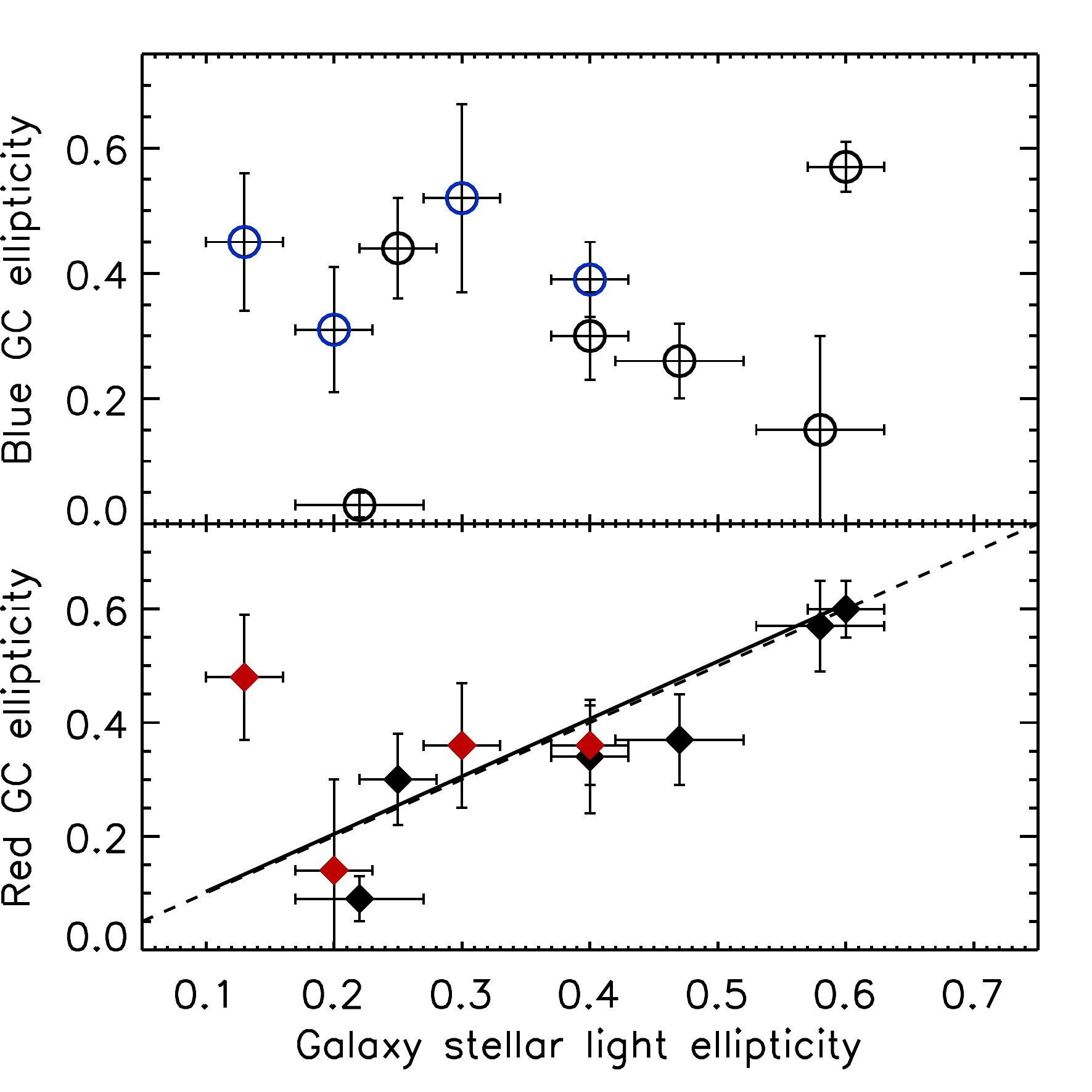} 
  \caption{GC ellipticity versus galaxy stellar light ellipticity. The top and the bottom panels show the relation between ellipticities of  blue GCs (open circles) and red GCs (filled diamonds) versus galaxy stellar light, respectively. Each panel includes an additional four galaxies (in red or blue colours) with the six galaxies (in black colour) from \citet{Kartha2014}. A linear fit to the red GCs (red and black diamonds) is drawn as a solid line and a one-to-one relation is shown as a dashed line. The red GC subpopulation confirms a one-to-one relation with galaxy stellar light, whereas only a weak relation for the blue GC subpopulation is present.} 
\label{ellip_plot}
\end{figure}

Figure \ref{ellip_plot} demonstrates the relation between GC subpopulation ellipticity and galaxy stellar light ellipticity for ten early-type galaxies. The plot is an updated version of figure 22 in \citet{Kartha2014}, with the addition of four galaxies (NGC 3607, NGC 3608, NGC 4406 and NGC 5813). Data for NGC 4406 and NGC 5813 are taken from \citet{Hargis2014}. For NGC 4406, the stellar, blue and red subpopulation ellipticities are  0.4 $\pm$ 0.03,  0.39 $\pm$ 0.06 and 0.36 $\pm$ 0.07, respectively. Similarly, the ellipticities for stellar, blue and red GC subpopulations of NGC 5813 are 0.3 $\pm$ 0.03, 0.52 $\pm$ 0.15 and 0.36 $\pm$ 0.11, respectively. With the addition of four galaxies, we observe a tight one-to-one relation between red GC ellipticity and galaxy stellar light. The relation is
\begin{equation}
\epsilon_\text{RGC}= [(1.0 \pm 0.1) \times \epsilon_\text{Stars}] + (-0.02 \pm 0.06)\\
\end{equation}
 The intrinsic scatter in the above relation is estimated as 0.10. The one-to-one relation signifies that the red subpopulations are affiliated with the stellar light of the parent galaxies \citep{Park2013}. In other words, both the red subpopulation and galaxy stellar light might have a similar origin. In contrast to \cite{Park2013}, we notice a poor association of blue GC subpopulation ellipticity with galaxy stellar light ellipticity. We explain this as a consequence of our wide-field imaging, as the ACSVCS data used by \citet{Park2013} does not reach far out enough to detect the whole  blue GC subpopulation for the most extended galaxies \citep{Peng2006}.  

In this small sample of ten galaxies, NGC 3607 shows the lowest ellipticity (nearly circular at $\epsilon$ = 0.13) for the galaxy stellar component. For NGC 3607, the ellipticities of both GC subpopulations show a deviation from the galaxy stellar light, although both are arranged along the photometric major axis of the galaxy.  So, NGC 3607 supports the idea that galaxies with low ellipticities might have randomly arranged GC subpopulations \citep{Wang2013}. The difference in spatial distribution of GC subpopulations from the galaxy stellar component suggests that a major fraction of both GC subpopulations might have formed separately from the galactic stars and later settled in the host galaxies. In the case of NGC 3608, both GC subpopulations show deviations from the galaxy stellar light in position angle.  In addition, the blue GC subpopulation shows a more elongated distribution than the red GC subpopulation. 

In addition to NGC 3608, three other galaxies - NGC 4365, NGC 4406 and NGC 5813 - also have blue GC subpopulations more elongated in shape than the red GC subpopulations. The elongated shape of the blue GC subpopulation suggests that it shows spatial distribution similarities with the red subpopulation that mostly follows the distribution of galaxy stellar component. If the distribution of blue GCs is not spherical, \citet{Wang2013} suggest that it may not have been built from accretions that were equally distributed in all directions. Instead they might formed through local filamentary structures in particular directions. This points out that directional dependent accretion or minor mergers might have occurred in these galaxies, altering the shape of blue GC subpopulations. 

In addition, these four elliptical galaxies are all slow rotators with kinematically distinct cores, KDCs, \citep{Emsellem2011,Krajnovic2011}. \citet{Naab2014} carried out hydrodynamical simulations to kinematically study the centres of early-type galaxies. They suggested that KDCs were generally formed in slow rotators that had experienced multiple gas-poor minor mergers. They proposed that their recent mass assembly histories are devoid of any major mergers and are expected to have older stellar populations. Few, if any GCs, are expected to have formed from such mergers. It is unclear whether blue GCs from the accreted galaxies would form a more elongated distribution than the host galaxy starlight as we observe.

In summary, the ellipticities of red GC subpopulations have a one-to-one relation with the galaxy stellar light ellipticities, whereas only a weak relation is seen for the blue GC subpopulation. Additionally, slowly rotating galaxies with a KDC have larger values for blue GC subpopulation ellipticities than their red GC counterparts. The elongated shape of the blue GC subpopulations may be due to recent minor mergers that were asymmetric in direction \citep{Tempel2015}.   

\subsection{GC metallicity gradients and galaxy stellar mass}

Colour gradients are important observational features for exploring the formation history of GC subpopulations and are clues to galaxy mass assembly. A negative colour gradient (GCs are redder at the centre of the galaxy than the outskirts) represents either the presence of younger (or more metal-rich) GCs at the galaxy centre or older (more metal-poor) GCs at the outskirts. As GCs are observed to be mostly old ($\sim$ 10 Gyr, \citealt{Strader2005,Dotter2010,Forbes2015}), the colour gradients are basically caused by metallicity gradients rather than age gradients. 

The observed gradients in GC subpopulations help discriminate between the different galaxy formation processes e.g., a negative gradient is predicted when the GCs are formed from a dissipative collapse \citep{Pipino2010}, while a gas-poor major merger will wash away any existing gradient \citep{Dimatteo2009}, a gas-rich major merger may remake a new gradient different from the original one \citep{Hopkins2009}, etc.  Also, minor mergers (accretions) can deposit GCs in the outskirts of galaxies \citep{Hirschmann2015,Pastorello2015} which will alter the existing gradient, perhaps resulting in an inner negative gradient and a flat outer gradient \citep{Oser2010, Forbes2011}.  

The first detection of a radial colour gradient in a GC system was by \citet{Geisler1996} in NGC 4472. With ground based data, GC colour gradients have been detected in other massive galaxies (NGC 4486: \citealt{Harris2009b}, NGC 1407: \citealt{Forbes2011}, NGC 4365: \citealt{Blom2012}), while only seen in a handful of intermediate mass galaxies (NGC 3115: \citealt{Arnold2011}, NGC 4564: \citealt{Hargis2014}) to date.

 In NGC 3607, another intermediate mass galaxy, the mean colours of both the blue and the red GC subpopulations reveal a significant colour gradient in the inner 6.5 arcmin (10 R$_e$). The colour gradient for the blue subpopulation is steeper than the red subpopulation. Within the total extent of the GC system (beyond 10 R$_e$), only the blue subpopulation has a significant colour gradient. We detect a significant colour gradient only for the blue GC subpopulation of NGC 3608.

\begin{table}
\centering
\caption{List of twelve galaxies observed with metallicity gradients for GC subpopulations. The metallicity gradients ($\Delta$[{\it Z/H}]) given below are obtained from the colour gradients. Galaxy name, logarithmic galaxy stellar mass, metallicity gradients for blue and red GC subpopulations with errors and the corresponding references (for colour gradient followed by the transformation equation used) are given.  }
\begin{tabular}{ccccc}
\hline
\multicolumn{1}{c}{Galaxy}&\multicolumn{1}{c}{log(M$_{\star}$)}&\multicolumn{2}{c}{Metallicity Gradient}&\multicolumn{1}{c}{Ref.}\\
NGC   & & Blue GCs & Red GCs & \\
&(M$_{\odot}$)& (dex dex$^{-1}$) & (dex dex$^{-1}$)&\\ 
\hline\hline
      1399 &   11.660 &   $-$0.12$\pm$0.05 &   $-$0.10$\pm$0.05  & ~1, 14\\
      1399 &   11.660 &   $-$0.21$\pm$0.04 &     --                           &  ~2, 11 \\
      1407 &   11.892 &   $-$0.22$\pm$0.04 &   $-$0.24$\pm$0.07 &  ~3, 10\\
      3115 &   11.249 &   $-$0.17$\pm$0.03 &   $-$0.24$\pm$0.06 &  ~4, 10  \\
      3115 &   11.239 &   $-$0.27$\pm$0.06 &   $-$0.11$\pm$0.10 &  ~5, ~5\\
      3607 &   11.677 &   $-$0.33$\pm$0.06 &   $-$0.16$\pm$0.07 &  ~6, 10\\                           
      3608 &   11.205 &   $-$0.25$\pm$0.05 &    --                           &  ~6, 10\\
      3923 &   11.796 &   $-$0.18$\pm$0.07 &   $-$0.17$\pm$0.08 &  ~5, ~5\\
      4278 &   11.290 &   $-$0.23$\pm$0.10 &   $-$0.23$\pm$0.12 &  ~7, 14  \\
      4365 &   11.843 &   $-$0.19$\pm$0.01 &   $-$0.22$\pm$0.03 &  ~8, 14   \\
      4472 &   12.046 &   $-$0.08$\pm$0.04 &   $-$0.10$\pm$0.05 &  ~1, 14\\
      4472 &   12.046 &   $-$0.13$\pm$0.03 &   $-$0.10$\pm$0.05 &  ~9, 11  \\
      4486 &   11.953 &   $-$0.12$\pm$0.02 &   $-$0.12$\pm$0.03 &  ~1, 14\\
      4486 &   11.953 &   $-$0.09$\pm$0.01 &   $-$0.12$\pm$0.01 &  10, 10\\
      4486 &   11.953 &   $-$0.17$\pm$0.07 &   $-$0.17$\pm$0.05 &  11, 11  \\
      4594 &   11.653 &   $-$0.17$\pm$0.04 &   $-$0.17$\pm$0.04 &  12, 12  \\
      4649 &   11.867 &   $-$0.00$\pm$0.04 &   $-$0.05$\pm$0.02 &   ~5, ~5\\
      4649 &   11.867 &   $-$0.21$\pm$0.05 &     --                          &  13, 14  \\
\hline
\end{tabular}
References: 1 - \citet{Liu2011}; 2 - \citet{Bassino2006a}; 3 -  \citet{Forbes2011}; 4 - \citet{Arnold2011}; 5 - \citet{Faifer2011}; 6 - This paper; 7 - \citet{Usher2013}; 8 - \citet{Blom2012}; 9 - \citet{Geisler1996}; 10 - \citet{Harris2009b}; 11 - \citet{Forte2012}; 12 - \citet{Hargis2014}; 13 - \citet{Strader2012}; 14 - \citet{Usher2013}
\label{CG_table}
\end{table}

\begin{figure}
\centering
 \includegraphics[scale=0.535]{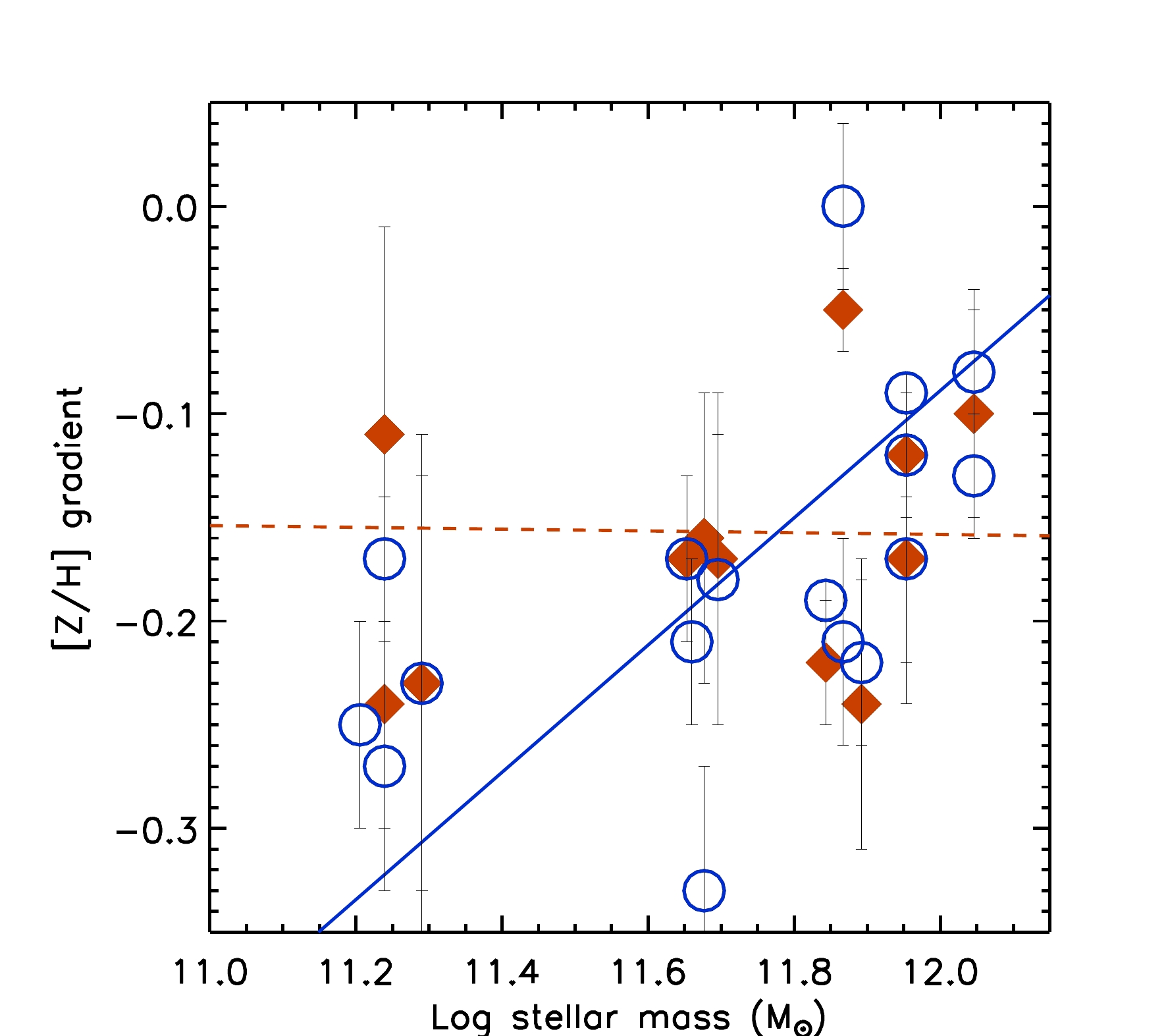}
\caption{Metallicity gradients of GC subpopulations versus galaxy stellar mass. The metallicity gradients (converted from colour gradients) for the blue (blue open circles) and the red (red filled diamonds) GC subpopulations for 12 galaxies are plotted against host galaxy stellar mass. The blue solid line and the red dashed line represent the linear fits to the metallicity gradients of blue and red GCs, respectively.  It is evident from the plot that for the blue GC subpopulation, the metallicity gradients become shallower with increasing galaxy stellar mass. Note that multiple measurements of metallicity gradients for the same galaxy are included. }
\label{CG_plot}
\end{figure}
 
 Table \ref{CG_table} compiles the list of galaxies in which colour gradients are detected for the GC subpopulations. The colour gradients are detected in different colour filters. For a uniform comparison, the different colour gradients are converted to metallicity gradients ($\Delta$[{\it Z/H}]) using colour-metallicity transformation equations (references are given in Table \ref{CG_table}).  The converted metallicity gradients are given in Table \ref{CG_table}. The list includes the gradients obtained only from wide-field imaging data and hence we exclude galaxies with single pointing {\it HST}/ACS imaging. This criterion excludes most galaxies from \citet{Liu2011}, for which only the central regions of target galaxies were covered. The sample of galaxies in the Table includes two galaxies from this paper plus another ten galaxies. The galaxy stellar masses are also tabulated. In order to calculate the galaxy stellar masses, we used their distance and  visual magnitude from NED and the mass to light ratios of \citet{Zepf1993}. In the sample of twelve galaxies, half have multiple measurements of their metallicity gradients. We include all measured GC gradients and their quoted uncertainties for the 12 galaxies. This comprises 18 measurements of blue GC gradients and 15 measurements of red GC gradients for 12 galaxies. 
 
In Table \ref{CG_table}, multiple measurements are given for five galaxies. For the same galaxy, the observed gradients are not always consistent among different works. For example, in the case of NGC 4649 the gradients for the blue subpopulation are $-$0.00 $\pm$ 0.04 \citep{Faifer2011} and $-$0.21 $\pm$ 0.05 \citep{Strader2011}. Both studies extended to 7 R$_e$. In another example, the gradient for the red subpopulation of NGC 3115 is quoted in \citet{Arnold2011} as $-$0.24$\pm$0.06, while \citet{Faifer2011} quoted $-$0.11$\pm$0.1.  But in the case of NGC 4472,  \citet{Geisler1996} and \citet{Liu2011} find consistent gradients for the red subpopulation.  

Figure \ref{CG_plot} shows the metallicity gradients for blue/red GC subpopulations versus the galaxy stellar mass.  We plot multiple measurements for individual galaxies. Linear fits are carried out separately for the blue and the red GCs with uncertainties estimated from the bootstrap technique. The technique uses the errors associated with individual gradients. Best fit relations are: 
\begin{eqnarray}
\Delta\text{[Z/H]}_\text{BGC} = [(0.31\pm0.08)\times \text{log(M}_{\star})] - (3.8\pm0.9)\\
\Delta\text{[Z/H]}_\text{RGC} = [(0.004\pm0.1)\times \text{log(M}_{\star})] - (0.1\pm1.0).
\end{eqnarray}
The galaxies in our sample have a mass range 11.0 < log(M$_{\star}$) < 12.0 M$_{\odot}$. From the above relations, we find that the metallicity gradient for the blue subpopulation has a significant correlation with stellar mass; the negative gradients flattens with increasing stellar mass. As more massive galaxies are expected to accrete more satellites \citep{Oser2010}, we expect more GC accretion to have taken place in these galaxies.  This addition of mostly blue GCs at different galactocentric radii may make the initial gradient of the blue GCs shallower. In addition, \citet{Hirschmann2015} found that gradients resulting from major mergers are shallower in more massive galaxies due to radial mixing of GCs. From the spectroscopic metallicities of GC subpopulations in twelve ETGs, \citet{Pastorello2015} observe a similar trend of decreasing metallicity gradient with increasing galaxy stellar mass.

For the red GC subpopulation, we are unable to find a significant relation between the metallicity gradient and galaxy stellar mass. In comparison with the blue GC subpopulations, the metallicity gradients for the red GC subpopulations have higher errors and also a lower number of data points. In Table \ref{CG_table}, the least significant gradient measurement is for the red subpopulation of NGC 3115 \citep{Faifer2011}.  Hence, we carried out another fitting for the red GCs without that measurement and the best fitted relation is
\begin{equation}
\Delta\text{[Z/H]}_\text{RGC} = [(0.07\pm0.05)\times \text{log(M}_{\star})] - (0.95\pm0.86).
\end{equation}
From the above relation, we infer that the gradients for the red GC subpopulation show a very weak dependence on galaxy stellar mass. 

The galaxy stellar mass is derived from the {\it M/L} ratios that are given in \citet{Zepf1993}. We appreciate that the \citet{Zepf1993} values are an approximation but have chosen to use them as this is the approach we took in \citet{Kartha2014}, which follows from the same approach as used by \citet{Rhode2010} and \citet{Spitler2008}. So in order to match the results with the above mentioned publications, we use the same method. \citet{BELL2003} derived the relationships to estimate the stellar mass from galaxy colours (see Appendix \ref{MLratio} for details). We used their relationship to derive  the galaxy stellar mass from ({\it B - V}) colour. We find that the equations 7 -- 9 are statistically unchanged when using \citet{BELL2003} to derive galaxy stellar masses.  

In summary, we suggest that the subsequent addition of GCs from minor mergers may weaken any pre-existing gradients in metallicity (from an early dissipative formation event) both for the blue and the red GC subpopulations.  

\subsection{Ratio of blue to red GCs with environment density and galaxy mass}

\begin{figure}
\centering
 \includegraphics[scale=0.475]{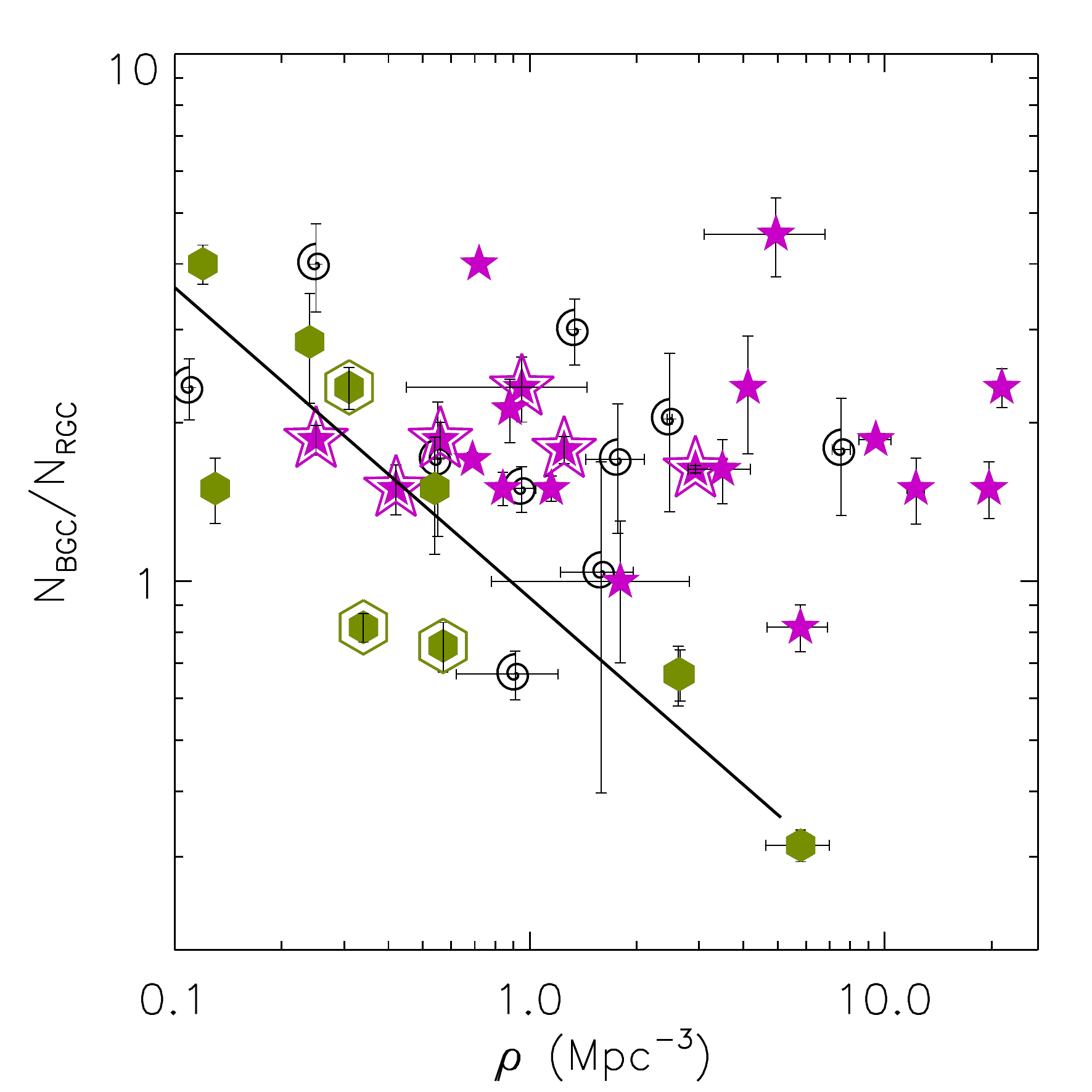} 
  \caption{Ratio of  blue to red GCs versus density of environment. Spirals, lenticular galaxies and elliptical galaxies are represented by spirals, hexagons and stars respectively. The galaxies from the SLUGGS survey are shown in double symbols. This is an updated version of figure 21 in \citet{Kartha2014} with the addition of NGC 3607 and NGC 3608. The position coordinates for NGC 3607 and NGC 3608 are respectively, (0.34, 0.79) and (0.56, 1.85). NGC 3607 is consistent with the correlation shown by lenticular galaxies.}
\label{rho_plot}
\end{figure}

\begin{figure}
\centering
 \includegraphics[scale=0.5]{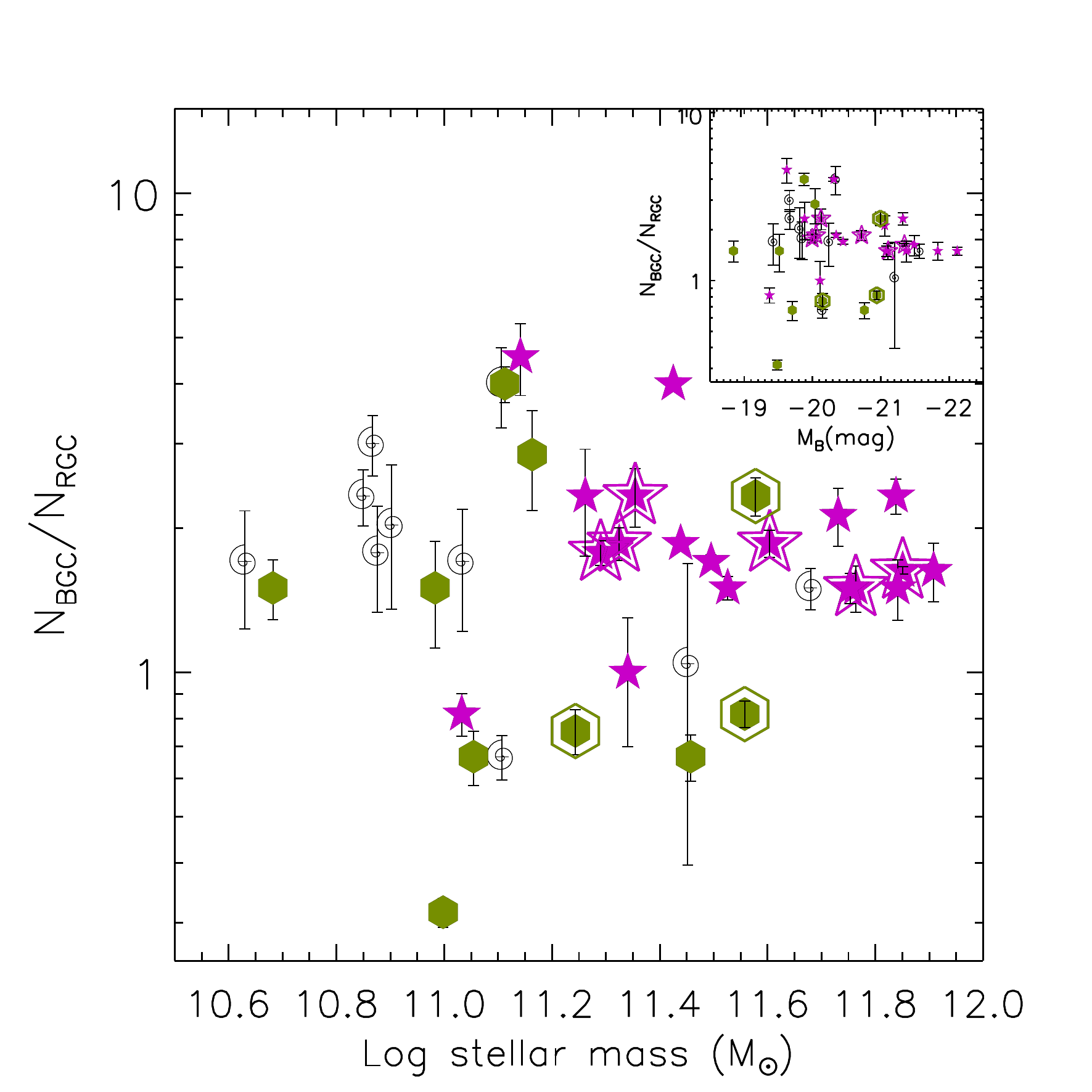} 
  \caption{Ratio of blue to red GCs versus galaxy stellar mass. The representation of symbols is same as in Figure \ref{rho_plot}. For all types of galaxies, there is no correlation between the ratio of blue to red GCs and galaxy stellar mass. They have a mean ratio of N$_\text{BGC}$/N$_\text{RGC}$  $\sim$ 1.7 for the total sample. An inset plot of ratio of blue to red GCs versus galaxy absolute {\it B}- band magnitude is given for the same sample. The mean ratio of blue to red GCs is $\sim$ 1.7 for the galaxy range $-$18.5 $<$ M$_{\text B}$ $<$ $-$22 mag. } 
\label{mass_plot}
\end{figure}

 Figure \ref{rho_plot} shows the ratio of blue to red GCs versus the environment density for a sample of 42 galaxies (two from this paper and  forty from \citealt{Kartha2014}). We note that the blue to red GC ratio is largely insensitive to any GC magnitude incompleteness. With this sample of galaxies, it is evident that neither spiral nor elliptical galaxies show any particular trend in the ratio of blue to red GCs with environment density. NGC 3608, with N$_\text{BGC}$/N$_\text{RGC}$ =  1.85 and density = 0.56 Mpc$^{-3}$  \citep{Tully1988}, is consistent with the other elliptical galaxies. 
  
\citet{Kartha2014} found that the fraction of blue to red GCs in lenticular galaxies decreases with local density of environment. This suggests that lenticular galaxies residing in high density environments accommodate a higher fraction of red GCs. NGC 3607 is a lenticular galaxy with a relatively high fraction of red GCs (N$_\text{BGC}$/N$_\text{RGC}$ =  0.79 and density = 0.34 Mpc$^{-3}$ from \citealt{Tully1988}).  The position of NGC 3607 is consistent with the trend of decreasing fraction of blue to red GCs with increasing  environment density.  
  
 In Figure \ref{mass_plot}, the ratio of blue to red GCs is plotted against the host galaxy stellar mass for the above sample of 42 galaxies. There is no obvious correlation between the ratio of blue to red GCs with the galaxy stellar mass. We divide the galaxies into three mass bins of size 0.5 M$_{\odot}$ and derive the mean value for the ratio of blue to red GCs. The mean ratio of blue to red GCs in the low (log(M$_{\star}$)< 11 M$_{\odot}$), intermediate (11 < log(M$_{\star}$) < 11.5 M$_{\odot}$) and high mass (log(M$_{\star}$) > 11.5 M$_{\odot}$) bins are respectively, 1.7 $\pm$ 0.8, 2.0 $\pm$ 1.2 and 1.6 $\pm$ 0.4. The mean ratio of blue to red GCs for the total sample of forty two galaxies is $\sim$ 1.76. 
  
 Using cosmological simulations, \citet{Bekki2008} investigated the structural, kinematical and chemical properties of GC systems in different Hubble type galaxies. They estimated the ratio of blue to red GCs, in the host galaxy luminosity range $-$ 14 < M$_{\text B}$ < $-$22 to vary from  $\sim$ 50 -- 0.25, with an average of 1.5. Using the ACSVCS, \citet{Peng2006} also investigated the ratio of blue to red GCs in a similar luminosity range and determined that the fraction varies from 5.6 to 0.67 percent from low to high luminosity galaxies, suggesting an average ratio of $\sim$ 1.5 blue to red GCs over the total luminosity range.
 
 There is a decreasing trend in the ratio of blue to red GCs with host galaxy luminosity, both observationally \citep{Peng2006} and theoretically \citep{Bekki2008}. We observe a nearly constant ratio of blue to red GCs in our sample of 42 galaxies because our luminosity range is much more restricted. As seen in the inset of Figure \ref{mass_plot}, our sample of forty two galaxies lie in the galaxy luminosity range  $-$18.5 $<$ M$_{\text B}$ $<$ $-$22 mag, whereas the faint end extends to M$_B$ = $-$14 mag for both \citet{Peng2006} and \citet{Bekki2008}.
 
 \subsection{Formation of GC systems}

\subsubsection{ NGC 3607 and NGC 3608}
In the Leo II group,  NGC 3607 is the massive central galaxy and has a red GC subpopulation fraction higher than the blue, while the neighbouring galaxy NGC 3608 is less massive and has a higher fraction of blue GCs. An overabundance of red GCs is observed along the minor axis of NGC 3607 (even after removing the GCs in the direction towards NGC 3608). From the azimuthal distribution of GCs of NGC 3608, it is found that both GC subpopulations are aligned in position angle and that angle is different from the position angle of the galaxy stellar light. These results (overabundance and misalignment) suggest a possible interaction between the galaxies in the group. 
  
Using {\it HST} data, \citet{Lauer2005} carried out an imaging study of 77 early-type galaxies, including NGC 3607. They detected an additional  gas disk settling in NGC 3607 perpendicular to the existing dusty disk. They commented that the dusty disk is in a transition phase merging with the gas disk. They explained this process as gas infalling directly onto the centre of NGC 3607 without disturbing the dusty disk and without any obvious features of interaction.  
 
 Later, \citet{Annibali2007} studied the stellar population properties of 66 early-type galaxies. They estimated the age, metallicity and alpha enhancement using the Lick indices with updated simple stellar population models (including the non-solar element abundance patterns). They estimated a very young age, 3.1 $\pm$ 0.5 Gyr, for NGC 3607 and suggested it had experienced a recent episode of star formation. \citet{Rickes2009} carried out long slit spectroscopy, out to galactocentric radii of 30.5 arcsec, and claimed that NGC 3607 has undergone a minimum of three star formation episodes with ages ranging from 1 to 13 Gyr.  The young age for the stellar population of NGC 3607 and the detection of a central gas disk indicate that NGC 3607 has experienced a recent star formation episode and the overabundance of red GCs may be due to GCs formed in that episode. 
 
 From the ATLAS$^{3D}$ survey, \citet{McDermid2015} estimated the mass-weighted ages for NGC 3605, NGC 3607 and NGC 3608 as 8.1 $\pm$ 0.8, 13.5 $\pm$ 0.7 and 13.0 $\pm$ 0.7 Gyr respectively. They utilised the spectra within 1R$_e$ to fit the single stellar population models and hence derive the mass-weighted ages, metallicity and star formation histories of 260 ETGs. Using the Lick indices, they estimated the age for NGC 3607 as 7.3 $\pm$ 1.3 Gyr that contradicts the young age determined by \citet{Annibali2007}. 
   
 \citet{Forbes2006a} carried out a multi-wavelength (X-ray, optical and H{\sc i} imaging) study of $\sim$ 60 galaxy groups, including the Leo II group. They investigated the evolutionary connections between different groups and the influence of group environment. In their study, they detected extended X-ray emission associated with the Leo II group but did not resolve individual galaxies. Recently, using Chandra X-ray data, \citet{Jang2014} observed X-ray emission from the central AGN in NGC 3607 and diffuse emission around NGC 3608. The detection of extended X-ray emissions confirms the presence of hot intergalactic gas. 
 
 The misalignment in the position angles of the GCs relative to the galaxy in NGC 3608 might be another sign of interaction with NGC 3607. Additionally, each galaxy shows an overabundance of GCs in the direct of the other, again suggesting a possible interaction between them. \citet{Jedrzejewski1988} proposed a close encounter between these two galaxies. They studied the absorption line kinematics for the stellar component  of NGC 3608 and found a change in direction of the rotation curve between the core and outside region. They proposed that the reversal might be due to an interaction with the nearby NGC 3607. 
 
 We conclude that our results also support a possible interaction between the two galaxies. To confirm this proposition, deep surface photometric and detailed kinematic studies are needed.

 \subsubsection{Formation scenarios}
  
As described in the introduction, three 'classic' formation scenarios were proposed to explain bimodality in globular cluster systems. In the major merger model \citep{Ashman1992}, the blue GCs already exist in the merging galaxies, while the red GCs form during the merging process. In the multi-phase collapse scenario \citep{Forbes1997}, the blue GCs are formed early, followed by a quiescent phase. After a few Gyrs, star formation is restarted with the formation of red GCs, which can be followed by accretion of additional blue GCs. According to \citet{Cote1998, Cote2000}, the red GCs are inherent to the parent galaxies and the blue GCs are purely accreted from dwarf galaxies. 

The three classic scenarios were explored in cosmological simulations which addressed a variety of GC system properties: structural and kinematical \citep{Bekki2005}, dynamical and chemical \citep{Bekki2008}, colour and metallicity bimodality \citep{Muratov2010,Tonini2013}, as well as physical relationships with the host galaxies \citep{Beasley2002}. Recently, \citet{Trenti2015} proposed another scenario for GC formation from the merging of multiple gas rich mini halos.

In all the classic formation scenarios, there is a strong association between red GC subpopulations and the parent galaxy. This relationship is established from different observations such as the strong relation between red GC peak colour and galaxy luminosity \citep{Peng2006,Strader2006,Faifer2011}, position angle arrangement of red GCs and the galaxy stellar component \citep{Wang2013}, connection between rotation velocity for red GCs and field stars \citep{Pota2013} etc. On the other hand, the association between blue GC subpopulations and parent galaxy stars is weak. \citet{Peacock2015} found that the blue GC subpopulations of NGC 3115 are consistent with the stellar halo in metallicity and spatial distributions. However, the origin of the blue GC subpopulation is quite controversial. \citet{Cote1998,Cote2000} and \citet{Tonini2013} proposed a dissipationless accretion origin whereas dissipational in-situ formation \citep{Forbes1997,Beasley2003} is suggested for the formation of blue GCs in the inner regions. This distinction in region (inner or outer) is mentioned since accretion of blue GCs to the galaxy outskirts in the later phase is also included in the multi-phase scenario \citep{Forbes1997}.  

\citet{Strader2004,Strader2005} investigated the feasibility of the above formation scenarios using observational data for massive elliptical galaxies. From the GC colour-galaxy luminosity relation and the age-metallicity relation, they proposed an in-situ plus accretion model for the formation of inner blue GCs which were then truncated by reionization, whereas the red GCs formed along with the bulk of field stars. They suggested that dwarf galaxies residing in overdense regions collapse before dwarfs in less dense regions, and then accrete more enriched gas from nearby star forming regions. These dwarf galaxies, along with their blue GCs, are later accreted into the halo of a massive galaxy forming part of the main system. This implies an in-situ+accretion origin for blue GCs. Hence, the origin of blue GCs in the inner regions could be due to one of three proposed processes, i.e. completely in-situ, fully accreted or in-situ+accretion.

In the following paragraphs, we try to differentiate between these three formation processes for blue GCs based on their global properties. In particular, we measure radial density, radial colour and azimuthal distributions in relation to their parent galaxies.  

From the azimuthal distribution of GC subpopulations, both blue and red GCs have a positional arrangement in common with the galaxy stellar light component \citep{Wang2013}. This suggests that the blue GC subpopulation and galaxy stellar component have similar evolutionary histories. For galaxies in which the blue GCs and stars accreted from satellite dwarfs, this similarity is expected \citep{Cote2001}. From the derived ellipticities it is seen that red GC subpopulations have a one-to-one relation with the galaxy stellar component whereas the relation is not tight for blue GC subpopulations (see Figure \ref{ellip_plot}).  If the galaxy has accreted its blue GCs recently, then a complete one-to-one correlation with host galaxy properties is not expected. \citet{Park2013} also investigated this relationship for 23 early-type galaxies using ACSVCS data and found an approximate one-to-one relation between blue GC ellipticity and the galaxy stellar component. As is well-known, the ACS field of view does not provide anywhere near complete coverage for massive nearby galaxies \citep{Peng2006}. That means a nearly one-to-one relation between inner blue GCs and galaxy stellar component suggests a common origin for both and hence, supports the in-situ formation scenario. 

Another diagnostic trend is the GC subpopulation peak colour versus galaxy luminosity. The peak colour of the red GC subpopulation gets redder with increasing galaxy luminosity. Perhaps a weaker correlation exists for the blue GC subpopulation. \citet{Liu2011} found that projection effects tend to flatten GC radial trends, particularly for the blue subpopulation because if its extended nature. Hence, the slope of the relation between the blue GC subpopulation peak colour and galaxy luminosity is reduced to half of the earlier value ($-$0.0126 $\pm$ 0.0025: \citealt{Peng2006}), making the relation between peak colour of the blue GC subpopulation and galaxy luminosity insignificant. This result weakens the idea that the formation of blue GCs is via in-situ processes.  

Radial colour gradients may also reveal the  origin of blue GCs. The colour gradients for blue GCs formed in-situ are expected to be steeper than for a subpopulation formed from in-situ+accretion or completely accreted processes. We expect this because the addition of GCs through accretion can dilute (in the case of in-situ+accretion) the existing colour gradient for the blue GC subpopulation. In the case of complete accretion, we assume zero colour gradient for the blue subpopulation.  Hence, to disentangle the origin of blue GCs, the steepness of the gradient needs to be quantified with large samples of galaxies where the colour gradients are measured with maximum accuracy. Our present work is limited by a small sample of 10 galaxies collected from the literature \citep{Geisler1996, Bassino2006a,Harris2009b, Forbes2011, Faifer2011,Arnold2011,Blom2012,Usher2013,Hargis2014} and two from this work. \citet{Liu2011} carried out an analysis of the colour gradients for 76 early-type galaxies using ACSVCS and ACS Fornax Cluster Survey (FCS). Even though the sample size is impressive, only three galaxies have more than one pointing and we have included them in the above sample. Hence, significant color gradients are detected in a total of 12 galaxies, five of which have multiple measurement. Gradient values are provided in Table \ref{CG_table}. 

Figure \ref{CG_plot} show this sample of GC metallicity gradients plotted against host galaxy stellar mass. The blue GC subpopulation shows a trend of decreasing gradient with increasing galaxy stellar mass. This implies that high mass galaxies have shallower gradients, whereas low mass (log(M$_\star$) $\sim$ 11.0 M$_\odot$) galaxies have steeper gradients. As the metallicity gradients show a dependency on galaxy stellar mass, both the GC subpopulations are expected to have some formational similarities with the galaxy stellar component. This means that a completely accreted origin \citep{Cote1998, Cote2000,Tonini2013} may not be the best scenario to explain the formation of blue GCs. Also, we notice that both GC subpopulation gradients show a dependence on galaxy stellar mass. Thus, a common, or in-situ origin \citep{Forbes1997,Beasley2003}, is probably involved in the formation of blue and red GC subpopulations \citep{Pastorello2015}. However, we note that large red (early-type) galaxies tend to preferentially accrete red satellite galaxies \citep{Hearin2014, Hudson2015}. Thus GC system metallicity gradients may also reflect the gradients of the accreted satellites, if they are preserved in the accretion process. In the in-situ+accretion formation scenario \citep{Strader2004,Strader2005} for the blue GCs, we expect the gradient to be shallower than for the blue GCs formed completely in-situ, but a reference scale is not yet established by models. 

To summarise, from the present study it is difficult to ascribe either a completely in-situ or an in-situ+accretion origin for the blue GC subpopulations. A homogeneous large sample with accurate GC properties is needed to address this issue in depth. 

\section{Conclusions}
We present wide-field imaging data from the Subaru telescope with which we carry out an investigation of the GC systems in the Leo II group to large galactocentric radii ($\sim$ 120 kpc). Using the multi-band wide-field images in {\it g}, {\it r} and {\it i} filters, we analysed the radial density, radial colour and azimuthal distributions of GC systems in the two brightest galaxies of the group, NGC 3607 and NGC 3608. Our study is complemented with spectroscopic data obtained from DEIMOS on the Keck II telescope. We present the main conclusions here.
\begin{enumerate}
	  \item The GC systems of NGC 3607 and NGC 3808 are found to have radial extents of 9.5 $\pm$ 0.6 arcmin (equivalent to 61 $\pm$ 5 kpc or $\sim$ 4.4 R$_e$) and 6.6 $\pm$ 0.8 arcmin (equivalent to 43 $\pm$ 5 kpc or $\sim$ 4.7 R$_e$), respectively. The derived values are in agreement with estimates obtained from the empirical relation between the effective radius of the GC system and galaxy stellar light given in \citet{Kartha2014}.
 	 \item The GC system colours of both galaxies are fitted with the GMM algorithm and we detect a bimodal distribution with confidence level greater than 99.99 percent. NGC 3607 is observed to have 45 $\pm$ 9 and 55 $\pm$ 8 percent of blue and red GC subpopulations, while for NGC 3608 the blue and red GC subpopulations contribute 65 $\pm$ 6 and 35 $\pm$ 6 percent to the total GC system.  
	 \item From the radial velocity measurements, we detect 81 GCs in the field of the Leo II group. We assign 46 and 35 GCs, respectively,  to NGC 3607 and NGC 3608. We estimate a mean velocity of 963 and 1220 km/s for NGC 3607 and NGC 3608, respectively. Also, the mean GC velocity dispersions for the respective galaxies are 167 and 147 km/s.
	  \item From the radial density distributions of the GC subpopulations of NGC 3607, the red subpopulation is more centrally located while the blue subpopulation is more extended.  Also, the effective radius of the red GC subpopulation (40 $\pm$ 29 arcsec) and the galaxy stellar light (39 arcsec) are in good agreement, compared to the blue subpopulation (95 $\pm$ 50 arcsec). 
	  \item For NGC 3608, the blue subpopulation is more extended in radius than the centrally concentrated red subpopulation. The red subpopulation distribution shows similarities with the galaxy surface brightness distribution. However, the effective radius of the red subpopulation (59 $\pm$ 40 arcsec) is larger than the galaxy stellar light (30 arcsec). 
 	 \item The azimuthal distribution of the NGC 3607 GC system reveals that both subpopulations are aligned along a position angle ($\sim$ 110 degrees), which is in reasonable agreement with the galaxy stellar light (125 degrees). However, the distribution of the GC system is more elliptical in comparison with the circular distribution of galaxy stellar light. The red subpopulation shows a more elliptical distribution when compared with the blue subpopulation.
    	\item For NGC 3608, the GCs are arranged along position angles that are different from the galaxy stellar population. Using two different methods of GC selection, the position angles for the total GC system are found to be along 104 $\pm$ 15 and 67 $\pm$ 7 degrees, while the galaxy major axis is at 82 degrees. One method of GC selection suggests that the GCs have an ellipticity = 0.20 $\pm$ 0.09, while the other shows an ellipticity of 0.39 $\pm$ 0.10. By comparison, the stellar light ellipticity is 0.20. In NGC 3608, the blue subpopulation has a more elliptical arrangement than the red subpopulation.
    	\item The total GC system, and both subpopulations of NGC 3607, become bluer in colour with increasing galactocentric radius; a significant metallicity gradient is observed for both subpopulations. We find that the blue subpopulation has a steeper gradient than the red subpopulation. We also detect a strong colour gradient only for the blue subpopulation of NGC 3608. The colour gradient for the blue subpopulation in NGC 3608 is steeper than that in NGC 3607.
	\end{enumerate}
	
	We compare different global properties of the GC systems and their parent galaxies. We reconfirm that the extent of the GC system is a function of galaxy size and the effective radius of a GC system is nearly 6 times the effective radius of parent galaxy. We obtain a one-to-one relation between the parent red GC ellipticities and galaxy stellar light ellipticities. Also, the blue GC ellipticities of slow rotators with kinematically decoupled cores are more elongated than their red GC subpopulation ellipticities. We propose that they might have experienced recent minor mergers from anisotropic  directions \citep{Tempel2015}. 
	
	From a sample of twelve galaxies, we investigate the relationship between the metallicity gradients and host galaxy stellar mass. We found that the gradients of both GC subpopulations become shallower with increasing stellar mass. The average ratio of blue to red GCs in galaxies in the mass range 11.0 $<$ log(M$_\star$) $<$ 12.0 M$_\odot$ is nearly 1.7. These findings agree with the predictions from the simulations of \citet{Bekki2008} and also with the findings from other observations \citep{Peng2006}. We also carried out a study to disentangle the formation of blue GC subpopulations (i.e. completely in-situ versus in-situ+accretion versus completely accreted), which have not given conclusive results and need to be followed up with a homogeneous, large sample.

\section*{Acknowledgments}
We thank the anonymous referee for his/her careful read- ing of the manuscript and the valuable feedbacks. The authors extend their gratitude to Jacob A. Arnold and Kristin A. Woodley regarding their help in observations. We thank Blesson Mathew and Nicola Pastorello for the careful reading of the manuscript. We also acknowledge the members of SAGES group, especially Christopher Usher, Vincenzo Pota and Joachim Janz, for the support and enlightening discussions.   SSK thanks the Swinburne University for the SUPRA fellowship. DAF thanks the ARC for support via DP-130100388. This work was supported by NSF grant AST-1211995. This paper was based in part on data collected at Subaru Telescope, which is operated by the National Astronomical Observatory of Japan. This paper uses data products produced by the OIR Telescope Data Center, supported by the Smithsonian Astrophysical Observatory.  Based on observations made with the NASA/ESA Hubble Space Telescope, and obtained from the Hubble Legacy Archive, which is a collaboration between the Space Telescope Science Institute (STScI/NASA), the Space Telescope European Coordinating Facility (ST-ECF/ESA) and the Canadian Astronomy Data Centre (CADC/NRC/CSA). Some of the data presented herein were obtained at the W. M. Keck Observatory, operated as a scientific partnership among the California Institute of Technology, the University of California and the National Aeronautics and Space Administration, and made possible by the generous financial support of the W. M. Keck Foundation. The authors wish to recognize and acknowledge the very significant cultural role and reverence that the summit of Mauna Kea has always had within the indigenous Hawaiian community. We are most fortunate to have the opportunity to conduct observations from this mountain. This research has made use of the NASA/IPAC Extragalactic Data base (NED) operated by the Jet Propulsion Laboratory, California Institute of Technology, under contract with the National Aeronautics and Space Administration. The analysis pipeline used to reduce the DEIMOS data was developed at UC Berkeley with support from NSF grant AST-0071048. We acknowledge the usage of the HyperLeda database (http://leda.univ-lyon1.fr).

\appendix

\section{List of spectroscopically confirmed objects around the Leo II group}
\label{appB}
Table \ref{vel_table} presents the photometric magnitudes {\it g}, {\it r} and {\it i} and the radial velocities for GCs, Galactic stars and background galaxies detected around NGC 3607 and NGC 3608 in the Leo II group.

\clearpage
\begin{deluxetable}{lcccccccccc}
\tablewidth{0pt}
\tabletypesize{\small}
\setlength{\tabcolsep}{5pt}
\tablecaption{Catalogue of objects detected around NGC 3607 and NGC 3608. The horizontal lines differentiate GCs of NGC 3607, GCs of NGC 3608, 7 ambiguous objects (classified into GCs and probable UCD - see Section \ref{kin_data}), Galactic stars and background galaxies. Column 1 represents the object ID with the galaxy name followed by the object classification such as GC, star and galaxy. Columns 2 and 3 present the position in Right Ascension and Declination (J2000). Columns 4 -- 9 present the Subaru/Suprime-Cam photometry in {\it g}, {\it r} and {\it i} filters and their respective uncertainties (given here are extinction corrected magnitudes). The heliocentric velocity and the respective uncertainty for each object is given in column 10 and 11.\label{vel_table}}
\tablehead{
 \colhead{ID} &
 \colhead{RA} &
 \colhead{Dec} &
 \colhead{{\it g}} &
 \colhead{{\it $\delta$g}} &
 \colhead{{\it r}} &
  \colhead{{\it $\delta$r}} &
 \colhead{{\it i}} &
 \colhead{{\it $\delta$i}} &
 \colhead{V$_{rad}$} &
 \colhead{$\delta$V}  \\
 \colhead{} &
 \colhead{(degree)} &
 \colhead{(degree)} &
 \colhead{(mag)} &
 \colhead{(mag)} &
 \colhead{(mag)} &
 \colhead{(mag)} &
 \colhead{(mag)} &
 \colhead{(mag)} &
 \colhead{(km/s)} &
 \colhead{(km/s)} \\
 \colhead{(1)} &
 \colhead{(2)} &
 \colhead{(3)} &
 \colhead{(4)} &
 \colhead{(5)} &
 \colhead{(6)} &
 \colhead{(7)} &
 \colhead{(8)} &
 \colhead{(9)} &
 \colhead{(10)} &
 \colhead{(11)} 
}
\startdata
\hline
  NGC3607\_GC1& 169.217229 & 18.003300  & 22.320 & 0.002 & 21.863 & 0.003 & 21.653 & 0.003  & ~958 & 14 \\ 
  NGC3607\_GC2&169.231425 & 18.022400 & 22.782 & 0.003 & 22.019 & 0.003 & 21.622 & 0.003  & ~924 & 9 \\  
  NGC3607\_GC3&169.298829 & 18.025774  & 22.706 & 0.004 & 22.228 & 0.003 & 22.027 & 0.004  & 1255 & 13 \\   
  NGC3607\_GC4& 169.268788 & 18.026516  & 22.646 & 0.003 & 21.898 & 0.002 & 21.533 & 0.003  & ~910 & 12 \\    
  NGC3607\_GC5&169.247763 & 18.029331  & 21.396 & 0.001 & 20.756 & 0.001 & 20.492 & 0.001  & ~904 & 5 \\     
  NGC3607\_GC6& 169.239792 & 18.027800  & 22.370 & 0.002 & 21.734 & 0.002 & 21.477 & 0.003  & ~825 & 15 \\ 
  NGC3607\_GC7&169.201883 & 18.032566  & 23.010 & 0.003 & 22.235 & 0.003 & 21.837 & 0.003  & ~961 & 17 \\
  NGC3607\_GC8& 169.251525 & 18.034105  & 22.218 & 0.002 & 21.596 & 0.002 & 21.361 & 0.002  & ~732  &13\\
 NGC3607\_GC9&  169.208850 & 18.035141 & 20.924 & 0.001 & 20.340 & 0.001 & 20.115 & 0.001  & ~950  &4 \\
 NGC3607\_GC10& 169.230150 & 18.037170& 22.053 & 0.002 & 21.321 & 0.002 & 21.027 & 0.002  & ~924 & 15 \\
 NGC3607\_GC11& 169.199067 & 18.037264 & 23.225 & 0.004 & 22.434 & 0.004 & 22.066 & 0.004  & 1136  &  15\\
 NGC3607\_GC12&  169.159492 & 18.033494  & 24.264 & 0.011 & 23.764 & 0.014 & 23.499 & 0.014  & 1052 &  12 \\
  NGC3607\_GC13& 169.236067 & 18.038172  & 22.161 & 0.002 & 21.525 & 0.002 & 21.279 & 0.002 & ~792  &  13 \\
 NGC3607\_GC14&  169.099217 & 18.039568  & 22.358 & 0.002 & 21.732 & 0.002 & 21.482 & 0.002  & 1048  &  15\\
 NGC3607\_GC15&  169.207533 & 18.043100 & 23.017 & 0.004 & 22.368 & 0.004 & 22.065 & 0.004  & ~764  & 17 \\
 NGC3607\_GC16&  169.213671 & 18.044300  & 23.698 & 0.005 & 23.089 & 0.006 & 22.770 & 0.007  & 1060  &  14 \\
  NGC3607\_GC17& 169.237692 & 18.047800 & 21.814 & 0.001 & 21.057 & 0.001 & 20.719 & 0.001  & ~920  &  12 \\
 NGC3607\_GC18&  169.238979 & 18.049000  & 23.307 & 0.007 & 22.484 & 0.007 & 21.903 & 0.006  & 1092  &  13 \\
  NGC3607\_GC19& 169.184475 & 18.049648 & 22.226 & 0.002 & 21.579 & 0.003 & 21.377 & 0.002  & ~966  &  13 \\
  NGC3607\_GC20& 169.194458 & 18.052082  & 22.753 & 0.003 & 22.098 & 0.003 & 21.713 & 0.004  & ~974  &  14 \\
  NGC3607\_GC21& 169.240721 & 18.054500  & 21.360 & 0.001 & 20.738 & 0.001 & 20.463 & 0.001  & ~598  & 9 \\
 NGC3607\_GC22&  169.220738 & 18.047815  & 21.831 & 0.001 & 21.313 & 0.001 & 21.092 & 0.002  & 1303  &  19 \\
  NGC3607\_GC23& 169.242683 & 18.055283 & 21.524 & 0.001 & 20.757 & 0.001 & 20.448 & 0.001  & ~840  &  9 \\
  NGC3607\_GC24& 169.265371 & 18.060005  & 22.899 & 0.003 & 22.149 & 0.003 & 21.811 & 0.003  & 1027 &  11 \\
 NGC3607\_GC25&  169.232196 & 18.058500 & 23.611 & 0.006 & 22.948 & 0.006 & 22.492 & 0.006  & ~987  &  13 \\
 NGC3607\_GC26&  169.271792 & 18.065065 & 23.112 & 0.004 & 22.420 & 0.004 & 22.110 & 0.004  & 1025  &  12 \\
 NGC3607\_GC27&  169.186392 & 18.065200  & 21.930 & 0.002 & 21.293 & 0.003 & 21.060 & 0.003  & ~848  &  16 \\
 NGC3607\_GC28&  169.219979 & 18.068300  & 21.646 & 0.001 & 20.972 & 0.001 & 20.655 & 0.001  & ~835  &  6 \\
 NGC3607\_GC29&  169.237104 & 18.074398  & 22.956 & 0.004 & 22.211 & 0.004 & 21.830 & 0.004  & ~958  &  16 \\
  NGC3607\_GC30& 169.213671 & 18.076300  & 24.242 & 0.009 & 23.568 & 0.009 & 23.240 & 0.01  & 1012  &  12 \\
 NGC3607\_GC31&  169.308429 & 18.075178  & 23.142 & 0.004 & 22.514 & 0.004 & 22.248 & 0.005  & ~949  &  14 \\
 NGC3607\_GC32&  169.229092 & 18.086800  & 22.753 & 0.003 & 22.003 & 0.003 & 21.650 & 0.003 & ~815 &  9 \\
  NGC3607\_GC33& 169.212542 & 18.087000  & 21.839 & 0.001 & 21.158 & 0.001 & 20.890 & 0.002  & ~985 &  7 \\
  NGC3607\_GC34& 169.160442 & 18.106018 & 22.294 & 0.002 & 21.748 & 0.002 & 21.521 & 0.003  & ~694  &  9 \\
 NGC3607\_GC35&  169.106842 & 18.140724 & 23.123 & 0.004 & 22.499 & 0.004 & 22.267 & 0.005  & 1138  &  18 \\
 NGC3607\_GC36&  169.376407 & 18.206726  & 22.467 & 0.002 & 21.941 & 0.003 & 21.741 & 0.003  & 1106  &  14 \\
 NGC3607\_GC37&  169.189700 & 18.011189  & 23.796 & 0.005 & 23.047 & 0.005 & 22.757 & 0.006  & ~723  &  17 \\
 NGC3607\_GC38&  169.211025 & 18.018753  & 22.581 & 0.002 & 22.030 & 0.003 & 21.810 & 0.003  & ~940  &  18 \\
 NGC3607\_GC39&  169.222354 & 18.045021 & 21.901 & 0.001 & 21.198 & 0.001 & 20.898 & 0.001  & 1145  &  23 \\
 NGC3607\_GC40&  169.203471 & 18.038366  & 22.804 & 0.004 & 22.081 & 0.004 & 21.699 & 0.004  & ~775  &  14 \\
 NGC3607\_GC41&  169.233100 & 18.040434 & 22.867 & 0.003 & 22.145 & 0.003 & 21.782 & 0.003  & ~873  &  26 \\
 NGC3607\_GC42&  169.242008 & 18.074684 & 24.662 & 0.013 & 24.083 & 0.016 & 23.961 & 0.018  & 1188  &  19 \\
 NGC3607\_GC43&  169.107099 & 18.244194 & 22.618 & 0.003 & 22.043 & 0.003 & 21.819 & 0.003  & 1279  &  16 \\
  \hline
  NGC3608\_GC1&169.231942 & 18.129700  & 25.547 & 0.021 & 24.951 & 0.023 & 24.783 & 0.028  & 1039  & 13 \\
  NGC3608\_GC2&169.246717 & 18.131090  & 22.657 & 0.003 & 22.041 & 0.003 & 21.806 & 0.003 & 1293  &  13 \\
  NGC3608\_GC3& 169.261129 & 18.135875  & 21.721 & 0.001 & 21.065 & 0.001 & 20.849 & 0.001  & 1091  &  6 \\
  NGC3608\_GC4&169.227192 & 18.138500 & 21.595 & 0.001 & 20.916 & 0.001 & 20.648 & 0.001  & 1272  &  8 \\
  NGC3608\_GC5&169.298158 & 18.138886 & 22.732 & 0.003 & 22.140 & 0.003 & 21.918 & 0.003  & 1242  & 15 \\
  NGC3608\_GC6&169.245833 & 18.146111  & 23.950 & 0.007 & 23.105 & 0.006 & 22.668 & 0.006  & 1176 &  29 \\
  NGC3608\_GC7&169.269692 & 18.139126 & 22.568 & 0.002 & 21.890 & 0.003 & 21.623 & 0.003 & 1458  &  9 \\
  NGC3608\_GC8&169.263733 & 18.142671  & 24.392 & 0.009 & 23.568 & 0.009 & 23.227 & 0.009 & 1141  &  15 \\
  NGC3608\_GC9&169.224058 & 18.145254  & 22.229 & 0.002 & 21.650 & 0.002 & 21.418 & 0.002 & 1493  &  14 \\
  NGC3608\_GC10&169.241266 & 18.14431  & 21.213 & 0.001 & 20.639 & 0.001 & 20.371 & 0.001  & 1203  &  12 \\
  NGC3608\_GC11&169.275825 & 18.151506  & 24.155 & 0.011 & 23.506 & 0.011 & 23.181 & 0.013  & 1467  &  19 \\
 NGC3608\_GC12& 169.298200 & 18.149405  & 22.834 & 0.003 & 22.243 & 0.003 & 22.044 & 0.004  & 1061  &  12 \\
  NGC3608\_GC13&169.222383 & 18.155500 & 22.851 & 0.003 & 22.260 & 0.003 & 22.029 & 0.004  & ~957  & 15 \\
  NGC3608\_GC14&169.243342 & 18.156600  & 22.166 & 0.002 & 21.581 & 0.002 & 21.258 & 0.002  & 1335  &  12 \\
  NGC3608\_GC15&169.229208 & 18.167700  & 23.038 & 0.003 & 22.479 & 0.003 & 22.272 & 0.004  & 1229  &  17 \\
  NGC3608\_GC16&169.324208 & 18.163569  & 22.591 & 0.002 & 21.995 & 0.003 & 21.763 & 0.003  & 1268   &  9 \\
  NGC3608\_GC17&169.260221 & 18.165211  & 22.266 & 0.002 & 21.670 & 0.002 & 21.487 & 0.002  & 1283 &  12 \\
  NGC3608\_GC18&169.273496 & 18.165411  & 23.057 & 0.005 & 22.558 & 0.005 & 22.414 & 0.005  & 1038  &  13 \\
  NGC3608\_GC19&169.292936 & 18.166916  & 21.915 & 0.002 & 21.332 & 0.002 & 21.130 & 0.002 & 1176  &  7 \\
  NGC3608\_GC20&169.229207 & 18.167725  & 22.623 & 0.002 & 21.976 & 0.002 & 21.741 & 0.003  & 1233  &  14 \\
  NGC3608\_GC21&169.252583 & 18.168700  & 22.874 & 0.003 & 22.248 & 0.004 & 22.035 & 0.004  & 1247  & 19 \\
  NGC3608\_GC22&169.320831 & 18.169830 & 22.265 & 0.002 & 21.576 & 0.002 & 21.289 & 0.002  & 1383  &  7 \\
  NGC3608\_GC23&169.337812 & 18.171459 & 22.885 & 0.003 & 22.267 & 0.004 & 22.031 & 0.004 & 1294  &  11 \\
  NGC3608\_GC24&169.220495 & 18.173113  & 23.444 & 0.006 & 22.925 & 0.005 & 22.703 & 0.006  & 1118  &  17 \\
  NGC3608\_GC25&169.259806 & 18.179447  & 22.292 & 0.003 & 21.759 & 0.003 & 21.582 & 0.003  & 1385  &  8 \\
  NGC3608\_GC26&169.237946 & 18.138157 & 23.728 & 0.005 & 22.946 & 0.005 & 22.619 & 0.005  & 1031  &  26 \\
  NGC3608\_GC27&169.271321 & 18.159868 & 23.265 & 0.004 & 22.567 & 0.004 & 22.266 & 0.004 & ~808  &  22 \\
  NGC3608\_GC28&169.228779 & 18.133675  & 22.476 & 0.002 & 21.936 & 0.002 & 21.736 & 0.003  & 1358  &  17 \\
  NGC3608\_GC29&169.266946 & 18.158815  & 23.398 & 0.004 & 22.629 & 0.004 & 22.304 & 0.004  & 1076  &  18 \\
 NGC3608\_GC30& 169.213571 & 18.159685  & 23.121 & 0.004 & 22.496 & 0.004 & 22.268 & 0.004  & 1238  &  25 \\
  NGC3608\_GC31&169.254055 & 18.160824 & 23.176 & 0.004 & 22.574 & 0.004 & 22.354 & 0.005 & 1328  &  12 \\
  NGC3608\_GC32&169.256616 & 18.169390  & 23.532 & 0.005 & 22.965 & 0.006 & 22.689 & 0.006  & 1180  &  19 \\
  \hline
  NGC3607\_GC44& 169.303350 & 18.082405  & 22.710 & 0.003 & 22.061 & 0.003 & 21.800 & 0.003  & 1318  &  11\\
  NGC3607\_GC45& 169.245171 & 18.095700  & 22.927 & 0.003 & 22.233 & 0.003 & 21.962 & 0.004  & 1089  &  14 \\
  NGC3607\_GC46& 169.184189 & 18.164358 & 22.793 & 0.003 & 21.974 & 0.003 & 21.599 & 0.003  & ~807  &  9 \\
  NGC3608\_GC33&169.203088 & 18.135864  & 22.218 & 0.002 & 21.563 & 0.002 & 21.314 & 0.002 & 1160  &  10 \\
  NGC3608\_GC34&169.217304 & 18.109472  & 22.905 & 0.004 & 22.333 & 0.004 & 22.084 & 0.005 & 1281  & 23 \\
  NGC3608\_GC35&169.192558 & 18.121163  & 23.079 & 0.004 & 22.484 & 0.004 & 22.250 & 0.005  & 1229  & 18 \\
  NGC3608\_ext1&169.197333 & 18.036472  & 22.384 & 0.0020 &21.893 & 0.0020 & 21.761 & 0.0030& 1822  & 22  \\
\hline
 NGC3608\_star1&169.181300       &18.000490&	22.633&	0.002&	22.119&	0.003&	21.950&	0.003&	 ~113 &	11\\
NGC3608\_star2	&169.189558&	18.004114&	21.235&	0.001&	20.761&	0.001&	20.625&	0.001& 	$-$106 &	 7\\
NGC3608\_star3	&169.223729&	18.025229&	25.096&	0.017&	24.464&	0.018&	24.073&	0.019& 	  ~~26 &	 4\\
NGC3608\_star4	&169.106475&	18.053684&	25.544&	0.032&	24.902&	0.031&	24.719&	0.039& 	  ~~90 &	10\\
NGC3608\_star5	&169.109129&	18.074558&	21.437&	0.001&	20.954&	0.001&	20.821&	0.002& 	 ~$-$37 &	10\\
NGC3608\_star6	&169.293779&	18.109633&	21.351&	0.001&	20.865&	0.001&	20.739&	0.001& 	 ~139 &	 7\\
NGC3608\_star7	&169.243571&	18.115063&	22.741&	0.003&	22.112&	0.003&	21.871&	0.004& 	 ~$-$59 &	17\\
NGC3608\_star8	&169.251296&	18.129112&	22.151&	0.002&	21.527&	0.002&	21.315&	0.002& 	 ~159 &	11\\
NGC3608\_star9	&169.198996&	18.147369&	21.484&	0.001&	20.985&	0.001&	20.849&	0.002& 	  ~~87 &	 9\\
NGC3608\_star10&169.148208&	18.176922&	25.108&	0.030&	24.433&	0.033&	24.134&	0.027& 	$-$107 &	16\\
NGC3608\_star11&169.389467&	18.250843&	22.643&	0.003&	22.093&	0.003&	21.938&	0.004& 	  ~~66 &	12\\
\hline
NGC3608\_gal1	&169.295992&	18.040196&	22.481&	0.003&	21.922&	0.003&	21.613&	0.003& 	- &	-\\
NGC3608\_gal2	&169.190583&	18.043064&	23.802&	0.006&	23.230&	0.006&	23.023&	0.008& 	- &	-\\
NGC3608\_gal3	&169.076800&	18.045994&	23.127&	0.004&	22.546&	0.004&	22.340&	0.005& 	- &	-\\
NGC3608\_gal4	&169.218121&	18.052868&	21.831&	0.001&	21.313&	0.001&	21.092&	0.002& 	- &	 -\\
NGC3608\_gal5	&169.111229&	18.095403&	25.878&	0.043&	25.324&	0.056&	24.929&	0.056& 	- &	-\\
NGC3608\_gal6	&169.125054&	18.108242&	22.672&	0.003&	22.205&	0.003&	22.070&	0.004&      -&	-\\
NGC3608\_gal7	&169.289267&	18.119137&	25.637&	0.025&	24.959&	0.032&	24.744&	0.040& 	-&	-\\
NGC3608\_gal8	&169.159358&	18.134090&	25.586&	0.027&	25.101&	0.032&	24.762&	0.037& 	- &	-\\
NGC3608\_gal9	&169.115850&	18.138969&	21.265&	0.001&	20.782&	0.001&	20.536&	0.002& 	- &	-\\
NGC3608\_gal10	&169.219142&	18.155857&	23.989&	0.008&	23.270&	0.008&	23.021&	0.009& 	- &	-\\
NGC3608\_gal11	&169.245833&	18.148889&	22.166&	0.002&	21.581&	0.002&	21.258&	0.002& 	- &	-\\
NGC3608\_gal12	&169.282651&	18.165146&	23.456&	0.005&	22.881&	0.006&	22.709&	0.008& 	-&	-\\
NGC3608\_gal13	&169.273496&	18.165411&	23.057&	0.005&	22.558&	0.005&	22.414&	0.005& 	-&	-\\
NGC3608\_gal14	&169.199783&	18.173775&	25.439&	0.041&	24.884&	0.042&	24.507&	0.038& 	- &	-\\
NGC3608\_gal15	&169.210004&	18.011913&	23.890&	0.007&	23.230&	0.008&	22.909&	0.009& 	-&	-\\
\hline
\enddata
\end{deluxetable}

\clearpage

\section{{\it M/L} ratio calculation using \citet{Bell2003} }
\label{MLratio}
 Relationships between stellar {\it M/L} values and various colours in SDSS and 2MASS passbands are given in \citet{BELL2003}.  They derived these relationships by fitting galaxy evolution models to a large sample of 22679 galaxies from the SDSS Early Data Release (\citealt{Stoughton2002}) and 2MASS extended source catalog (\citealt{Jarrett2000}). To estimate the stellar mass for our sample of 42 galaxies, we utilize the relationship between {\it M/L} ratio and ({\it B $-$ V}) colour which is given below.
\begin{equation}
\text{log}_{10}(M/L) = -0.628 + (1.305 \times (B-V))
\end{equation}

We find that the \citet{BELL2003} M/L ratios are about a factor of $\sim$ 2 times lower for ellipticals and a factor of $\sim$ 1.5 times lower for lenticulars than \citet{Zepf1993} values. This affects the X-axes of Figures 18 and 20. Hence, we fit the trends in Figure 18 after incorporating the stellar mass from \citet{BELL2003}. The fits are given below which can be compared to Equations 7 -- 9. 
\begin{eqnarray}
\Delta\text{[Z/H]}_\text{BGC} = [(0.33\pm0.07)\times \text{log(M}_{\star})] - (3.9\pm0.8)~~~~\\
\Delta\text{[Z/H]}_\text{RGC} = [(0.0007\pm0.06)\times \text{log(M}_{\star})] - (0.15\pm0.9)\\
\Delta\text{[Z/H]}_\text{RGC} = [(0.07\pm0.05)\times \text{log(M}_{\star})] - (0.89\pm0.75)~
\end{eqnarray} 
We find that even if the stellar mass varies between \citet{Zepf1993} and \citet{BELL2003}, the relationships shown by blue and red GCs with metallicity remains statistically the same. This also implies that our results remain unchanged between different {\it M/L} ratio estimations.

\bibliographystyle{mn2e}

\bibliography{reference}

\end{document}